\documentclass{jfm}

\usepackage{amsmath}
\usepackage{graphicx}
\usepackage{newtxtext}
\usepackage{newtxmath}
\usepackage{bm}
\usepackage{xcolor}
\newcommand{\rev}[1]{{\color{black}#1}}
\usepackage{booktabs}
\usepackage{longtable}
\usepackage{pifont}

\usepackage{tikz}
\usetikzlibrary{arrows.meta,positioning,fit,backgrounds}
\usepackage{natbib}
\usepackage{hyperref}
\hypersetup{
    colorlinks = true,
    urlcolor   = blue,
    citecolor  = black,
}

\graphicspath{{figures/}}

\newcommand{\bU}{\mathbf{U}}

\newcommand{\tij}{\boldsymbol{\tau}}
\newcommand{\Smat}{\mathbf{S}}
\newcommand{\Wmat}{\mathbf{\Omega}}
\newcommand{\Tw}{\mathbf{T}_2}        
\newcommand{\Tz}{\mathbf{T}_0}        
\newcommand{\Gz}{g_0}                 
\newcommand{\tzz}{\tau_{zz}}          
\newcommand{\nut}{\nu_t}
\newcommand{\lmix}{\ell}              
\newcommand{\Cmu}{C_\mu}              
\newcommand{\kvk}{\kappa}             

\title{Generalizable turbulence closures across bluff-body shapes by PINN-based solver-agnostic training}

\shorttitle{Solver-agnostic neural turbulence closures}
\shortauthor{Z. Zhang, T. K\"aufer, L. Ronglan, M. S. Triantafyllou, and G. E. Karniadakis}

\author{
  Zhen Zhang\aff{1},
  Theo K\"aufer\aff{2},
  Louise Ronglan\aff{2},
  Michael S. Triantafyllou\aff{2},
  George Em Karniadakis\aff{1}\footnotemark[1]
}
\affiliation{\aff{1}Division of Applied Mathematics, Brown University
\\ \aff{2}Department of Mechanical Engineering, Massachusetts Institute of Technology
}
\begin{document}
\maketitle
\begingroup
\renewcommand\thefootnote{\fnsymbol{footnote}} 
\footnotetext[1]{Email address for correspondence: george\_karniadakis@brown.edu}
\endgroup

\begin{abstract}

\rev{Data-driven turbulence closures are commonly calibrated by embedding a CFD solver within an inverse procedure that requires a converged flow solution at every optimization iteration, which is fragile and costly. We instead formulate closure identification as a PDE-constrained, data-driven inverse problem within a physics-informed neural network (PINN), and subsequently stabilize the inferred closure for deployment in a numerical solver. This separation eliminates the training cost of repeated CFD solves and associated adjoint or ensemble evaluations, providing substantial flexibility to investigate alternative closure forms. Using this flexibility, we develop four closures:} three model the Reynolds stress on a realizable tensor basis---a local map, a non-local model transporting the turbulent kinetic energy and recovering the out-of-plane normal stress, and the same with a learned length scale $\ell$---and a fourth models the Reynolds force $\mathbf{F} = -\nabla\!\cdot\boldsymbol{\tau}$ directly, free of the realizability constraint. All four are trained across six two-dimensional bluff-body wakes at $\Rey = 10^4$ and deployed frozen in a standard finite-element solver, stabilized \rev{primarily by input-gradient smoothing}. Under a strict leave-one-shape-out (LOSO) protocol, all four \rev{reduce the field errors of a steady SST $k$--$\omega$ baseline, most strongly on the mean velocity}. The learned-length-scale closure is most accurate on the stress fields, while the force model generalizes best on the mean velocity and drag (LOSO drag error ${\sim}7\%$). The closures also train efficiently on Particle Image Velocimetry data, enabling geometries intractable for DNS.

\end{abstract}


\section{Introduction}\label{sec:intro}

Reynolds-averaged Navier--Stokes (RANS) simulation remains the workhorse of industrial
computational fluid dynamics (CFD), and its predictive accuracy is governed almost entirely by the
turbulence closure---the model for the unknown Reynolds stress $\tij=\langle u_i' u_j'\rangle$.
Classical closures are accurate only within the regime for which they were calibrated, and the
last decade has seen sustained effort to learn closures from high-fidelity data
\citep{duraisamy2019,brunton2020}.
\rev{Two limitations motivate the present approach. Training with a CFD solver in the
optimization loop is computationally costly and numerically fragile, because intermediate
closure models must support convergent flow solutions. Moreover, coupling the training
procedure to a particular solver and its closure implementation restricts the flexibility
with which alternative closure forms can be investigated. Purely a-priori supervised
training avoids these repeated flow solves, but does not directly control the a-posteriori
error arising when the learned closure is coupled to the governing equations. It also
typically requires paired observations of the closure inputs and outputs, including local
velocity gradients and turbulence quantities, which are rarely available in full from
experiments. This work therefore addresses  the following four questions: 
\begin{enumerate}
    \item How can a PDE-constrained
inverse formulation identify closures from incomplete observations without repeated CFD
solves during training?
\item How can the training framework accommodate alternative closure
forms and modeling targets?
\item What numerical treatment enables stable deployment of the
inferred closure in a flow solver?
\item How well do the resulting closures generalize to
geometries excluded from training?
\end{enumerate}}

\rev{\emph{Solver-in-the-loop training and its limitations.}}
Many data-driven approaches solve the inverse closure problem by embedding a CFD solver
within an optimization loop. Field-inversion-and-machine-learning infers spatial corrections
through an adjoint calculation and subsequently regresses a closure map
\citep{parish2016,singh2017}; ensemble methods update closure parameters using multiple
forward simulations \citep{zhang2022enkf}; and CFD-in-the-loop symbolic approaches evaluate
candidate expressions through their deployed flow solutions \citep{weatheritt2016,zhao2020}.
Neural closures can also be optimized directly through the governing-equation adjoint,
with successful applications including large-eddy simulations of bluff-body flows
\citep{sirignano2020dpm,sirignano2023dlles}. \rev{These approaches incur the cost of repeated
forward and sensitivity calculations, but their difficulty extends beyond computational
expense. The nonlinear Navier--Stokes equations and the state dependence of the neural
closure form a coupled system whose local well-posedness and numerical stability are
generally difficult to ensure \citep{wu2019,brener2021,zhang2026tangent}.
A closure update changes both the constitutive response
and the conditioning of the flow problem; consequently, an intermediate model may prevent
the embedded solver from converging before its agreement with observations can be assessed.
Requiring stable flow solutions throughout training therefore imposes an additional
constraint on the search for an accurate closure, particularly from randomly initialized
network parameters. Flexibility is also limited by the solver's discretization and
implementation of the closure terms: changing the modelling target from Reynolds stress
to Reynolds force, for example, can require substantial modifications to the flow solver
and its sensitivity calculations.}

\rev{\emph{PINNs for inverse identification of turbulence closures.}
These considerations motivate separating model
identification from stabilization of the deployed equations. A physics-informed neural
network (PINN) provides a flexible, PDE-constrained formulation of the inverse problem,
in which the mean flow and closure are inferred jointly from observations and the
governing-equation residuals \citep{raissi2019}. Training thus avoids requiring a converged
CFD solution for every intermediate closure, while numerical stabilization can be developed
and assessed for the inferred model at deployment.} PINN optimization itself remains
challenging for separated turbulent flows, but recent advances in preconditioned optimizers
\citep{gupta2018shampoo,vyas2024soap}, adaptive gradient-norm loss balancing
\citep{wang2021grad}, and pseudo-time relaxation \citep{cao2023tsonn,wang2026pseudo}
have improved training robustness and make this formulation increasingly practical for
closure identification. Earlier PINN--RANS studies primarily reconstructed the mean flow
and stresses for individual cases \citep{eivazi2022,patel2024}. \rev{Our previous work instead
used a PINN to identify a transferable Reynolds-force closure for cylinder flows, followed
by solver-in-the-loop retraining in OpenFOAM \citep{zhang2026}; this additional stage was
needed in the absence of a sufficiently general stabilization procedure for direct
deployment of the inferred closure. The present work extends that approach by combining
PINN-based closure identification with a separate numerical stabilization strategy, so
that the learned closure can be deployed without solver-in-the-loop retraining, and by
extending the assessment from a cylinder to distinct bluff-body geometries.}

\emph{A-posteriori stability.}
Finally, a closure is only useful if it can be substituted into a solver. Deploying a frozen
data-driven stress as an explicit source term is a recognized failure mode rather than a
hypothetical risk: \citet{wu2019} showed the RANS operator is ill-conditioned with respect to the
Reynolds stress, so sub-$0.5\%$ stress errors can produce velocity errors up to $\sim\!35\%$. This
is the mechanism behind the well-documented gap between a-priori accuracy against stored data and
a-posteriori accuracy in a live solve \citep{brunton2020,duraisamy2021}; conditioning persists even
under implicit eddy-viscosity treatment unless velocity information is incorporated
\citep{brener2021}, motivating model-consistent training that re-embeds a solver in the loop
\citep{michelen2021}.
\rev{The state dependence of a neural closure introduces a further issue: local
solvability and sensitivity depend on the Jacobian of the coupled mean-flow and closure
operators, rather than on the closure error alone. Adding a neural closure can compromise the well-posedness of the original PDE
and introduce multiple, potentially non-physical solutions \citep{zhang2026tangent}.}
Clearing the bar of stable a-posteriori deployment in a standard,
non-PINN solver is thus a genuine challenge. We address it \rev{primarily with input-gradient
smoothing, supported by a Lipschitz constraint on the learned closure}.

\emph{Closure form: Reynolds stress or Reynolds force.}
\rev{Tensor-basis representations express Reynolds-stress anisotropy in terms of
strain-rate and rotation-rate tensors} \citep{pope1975}, \rev{providing a common
foundation for algebraic Reynolds-stress models, nonlinear eddy-viscosity models,
and tensor-basis neural networks}
\citep{gatski1993,wallin2000,craft1996,ling2016}.
\rev{Realizability imposes additional constraints on the predicted stress tensor}
\citep{schumann1977,lumley1978}.
\rev{Beyond local constitutive relations, transport equations for turbulence quantities
introduce dependence on upstream evolution; the Prandtl--Kolmogorov relation connects
eddy viscosity to turbulent kinetic energy and a characteristic length scale}
\citep{wolfshtein1969velocity}.
\rev{An alternative modelling target is the Reynolds force,
$\mathbf{F}=-\nabla\!\cdot\tij$, which enters the mean momentum equation directly.
Learning this vector instead of the full stress tensor has been explored by}
\citet{cruz2019}, \rev{while conditioning analyses highlight the importance of the
closure representation and its coupling to the mean flow} \citep{brener2021}.

\emph{Generalization across geometries.}
Generalization in Reynolds number on a fixed geometry is addressed in our previous 
work \citep{zhang2026}; generalization
across geometries, where the input features leave the training manifold, is widely regarded
as the central open difficulty for data-driven closures \citep{duraisamy2019}. Cross-geometry or
leave-one-shape-out (LOSO) validation is a recognized and demanding test: tensor-basis neural networks and
random forests have been cross-applied across families such as ducts, periodic hills and
backward-facing steps \citep{ling2016,kaandorp2020}, curated multi-geometry datasets now support
consistent benchmarking \citep{mcconkey2021}, and \citet{huijing2021} cross-applied trained closures between three-dimensional bluff bodies. Documented gains are real but bounded,
typically improvement over a baseline within a family of related shapes. We therefore adopt a
strict leave-one-shape-out protocol across six distinct bluff-body wakes as a high-bar test of
geometry generalization.

\begin{figure}
    \centering
    \includegraphics[width=0.8\linewidth]{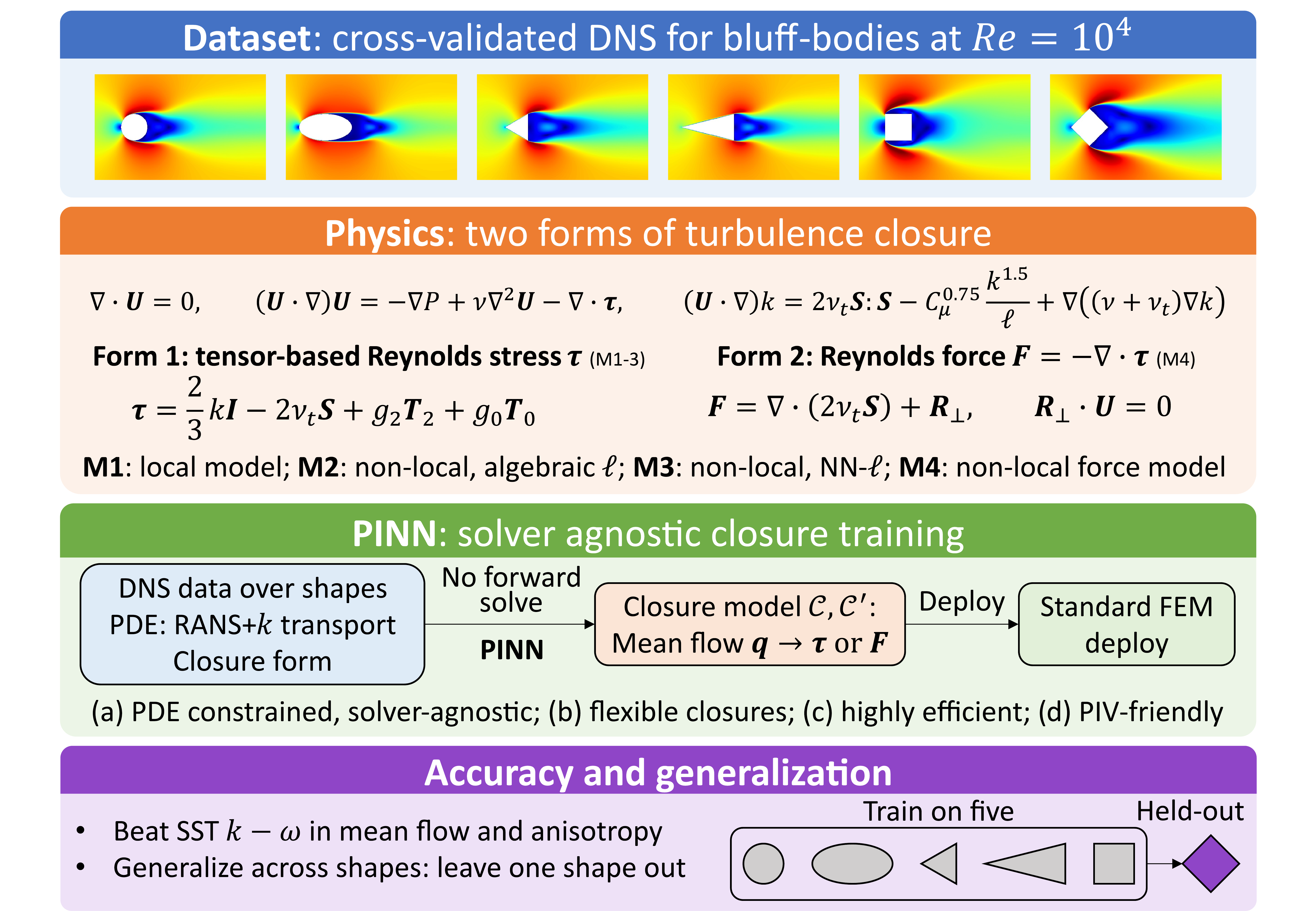}
    \caption{Overview of the solver-agnostic PINN closure framework.}
    \label{fig:sketch}
\end{figure}

\emph{Contributions.} Figure~\ref{fig:sketch} summarizes the content of this paper. In summary,
we (i) use a PINN as a solver-agnostic, PDE-constrained, differentiable
trainer of RANS closures---imposing the governing residual by automatic differentiation with no
forward solve in the loop, so that only the final closure need be solver-stable, an arbitrary
neural closure is admitted without deriving its adjoint, and the iterative cost of adjoint or
ensemble methods is avoided;
(ii) within this framework, develop four closures on equal footing that
model either the Reynolds stress or the Reynolds force;
(iii) deploy the frozen closures stably in a standard
finite-element RANS solver \rev{primarily through input-gradient smoothing};
and (iv) test all four closures under a strict leave-one-shape-out protocol across six
distinct bluff bodies, and show that the learned-length-scale closure (M3) is the most accurate on
the stress field while the force model (M4) generalizes best on the mean velocity and recovers the
drag \rev{with a mean error of $\sim$7\%}, all four outperforming the SST $k$--$\omega$ baseline \rev{on every field quantity in-sample and
out-of-sample, by a wide margin on the mean velocity}.

The remainder of the paper is organized as follows. Section~\ref{sec:dataset} describes the six
bluff-body wakes and the DNS, PIV, and SST $k$--$\omega$ reference data. Section~\ref{sec:method}
develops the methodology: \rev{the governing equations (\S\ref{sec:method-form}), the four closure forms and their
solver-agnostic PINN training (\S\ref{sec:method-pinn})}, and the frozen finite-element deployment together with its
stabilization (\S\ref{sec:method-solver}). Section~\ref{sec:results} reports the results---the
PDE-consistency that distinguishes the trained closure from an a-priori fit
(\S\ref{sec:results-apriori}), the stabilization ablation (\S\ref{sec:results-stab}), training
directly from patched PIV data (\S\ref{sec:results-piv}), and the in-sample
(\S\ref{sec:results-insample}) and leave-one-shape-out (\S\ref{sec:results-loso}) accuracy of all
four closures. Section~\ref{sec:conclusion} concludes. \rev{The appendices compare the
PIV and DNS mean fields for the diamond and long triangle (appendix~\ref{app:piv}), detail the
reconstruction and parameterization of the four closures (appendix~\ref{app:closure-training}),
give the PINN setup, the supervised-PINN fields and an ablation of the training recipe
(appendix~\ref{app:pinn}), quantify the sensitivity of the deployed solution to the
input-filter constant (appendix~\ref{app:filter-sensitivity}), and collect the in-sample and
leave-one-shape-out field comparisons for the remaining shapes (appendix~\ref{app:fields}).}

\section{Dataset and flow configuration}\label{sec:dataset}

\rev{We consider two-dimensional mean turbulent wakes behind six bluff bodies at
$\Rey=U_\infty D/\nu=10^4$, with $U_\infty=D=1$. The geometries span smooth and
sharp-edged boundaries and different orientations, providing distinct separation
patterns and wake structures for assessing transfer across shapes. Their dimensions
and the common reference-data configurations are summarized in
table~\ref{tab:dataset-setup}; the mean wakes are shown in figure~\ref{fig:sketch}.}

\begingroup
\color{black}
\renewcommand{\rev}[1]{{\leavevmode\color{black}#1}}
\small
\setlength{\tabcolsep}{3pt}
\renewcommand{\arraystretch}{1.13}
\begin{longtable}{@{}p{0.19\linewidth}p{0.25\linewidth}p{0.51\linewidth}@{}}
\caption{Geometries and reference-data configurations.}\label{tab:dataset-setup}\\
\toprule
\endfirsthead
\multicolumn{3}{l}{\tablename\ \thetable\ (continued)}\\
\toprule
\endhead
\bottomrule
\endfoot
\multicolumn{3}{@{}l}{\textbf{1. Body geometries}}\\*
\multicolumn{3}{@{}p{0.98\linewidth}@{}}{%
\begin{tabular}[t]{@{}p{0.18\linewidth}p{0.28\linewidth}p{0.19\linewidth}p{0.28\linewidth}@{}}
\textbf{Shape} & \textbf{Dimensions} & \textbf{Shape} & \textbf{Dimensions}\\
Circle & Diameter $D$ & Ellipse & Length $2D$; height $D$\\
Square & Side $D$ & Diamond & Side $D$; frontal diagonal $\sqrt{2}D$\\
Long triangle & Base $D$; length $2D$ & Equilateral triangle & Side $D$\\
\end{tabular}}\\
\midrule
\multicolumn{3}{@{}l}{\textbf{2. DNS: common setup for all six shapes}}\\*
\textbf{Numerical setting} & \textbf{Method} & \textbf{Specification}\\*
Solver and domain & NekRS; three-dimensional &
\rev{$y/D\in[-20,20]$. Circle and both triangles: $x/D\in[-12,20]$, $L_z/D=2\pi$;
other shapes: $x/D\in[-12,30]$, $L_z/D=1.5\pi$. Spanwise-periodic boundaries.}\\
Spatial resolution & \rev{Spectral elements; $N=9$ for circle, $N=7$ otherwise} &
\rev{Approximately $1.5$--$2.5\times10^5$ hexahedral elements and
$0.8$--$1.5\times10^8$ grid points;} body resolution $y^+<1$; near-isotropic wake cells for
$3<x/D<10$, $-3<y/D<3$.\\
Time integration & BDF2/EXT2 & Dealiased, over-integrated advection; variable time step at a fixed Courant number.\\
Statistics & Time and spanwise averages & Statistics collected after the initial transient:
$U,V,P,\tau_{xx},\tau_{xy},\tau_{yy},\tau_{zz}$.\\
\rev{Averaging duration} & \rev{$T_{\mathrm{avg}}U_\infty/D$; shedding periods $T_{\mathrm{avg}}U_\infty St/D$} &
\rev{circle: $99$ ($19.4$ periods); ellipse: $400$ ($110.0$ periods); square: $160$ ($20.6$ periods); diamond: $160$ ($19.2$ periods); long triangle: $150$ ($35.4$ periods); equilateral triangle: $149.9$ ($28.3$ periods).}\\
\midrule
\multicolumn{3}{@{}l}{\textbf{3. PIV: circle, diamond, and long triangle only}}\\*
\textbf{Measurement} & \textbf{Method or extent} & \textbf{Acquisition and processing}\\*
Acquisition & Planar, time-resolved PIV & \rev{6380 snapshots per case at 500 Hz;}
normalization by body dimension and towing speed.\\
\rev{Averaging duration} & \rev{$12.76$ s per case} &
\rev{Approximately 37.4 (circle), 21.8 (diamond), 46.7 (long triangle) vortex-shedding periods.}\\
Processing & DAVIS 11 (LaVision) & Final interrogation windows $32\times32$ pixels;
$75\%$ overlap; approximately $450\times350$ vectors with spacing $0.014D$.\\
Coverage and fields & $x/D\in[-0.2,6.1]$; $y/D\in[-2.5,2.5]$ &
In-plane velocity and Reynolds stresses; pressure and out-of-plane velocity are not measured.\\
\midrule
\multicolumn{3}{@{}l}{\textbf{4. OpenFOAM RANS reference: all six shapes}}\\*
\textbf{Turbulence model} & \textbf{Domain} & \textbf{Mesh resolution}\\*
SST $k$--$\omega$ & \rev{$x/D\in[-10,15]$, $y/D\in[0,10]$ (half-domain)} &
\rev{4930--5895 finite-volume cells; one cell in the spanwise direction.}\\
\end{longtable}
\endgroup

\rev{Reference statistics are obtained from three-dimensional DNS using NekRS}
\citep{fischer2022nekrs}, \rev{following the configuration of} \citet{zhang2026}.
\rev{Time and spanwise averaging provide the mean velocity, pressure, and Reynolds
stresses, including the spanwise normal stress needed to determine the full turbulent
kinetic energy. For field comparisons, DNS and numerical predictions are evaluated
on the common Cartesian window $x/D\in[-3,8]$, $y/D\in[-3,3]$.}

\begin{figure}
  \centering
  \includegraphics[width=0.8\textwidth]{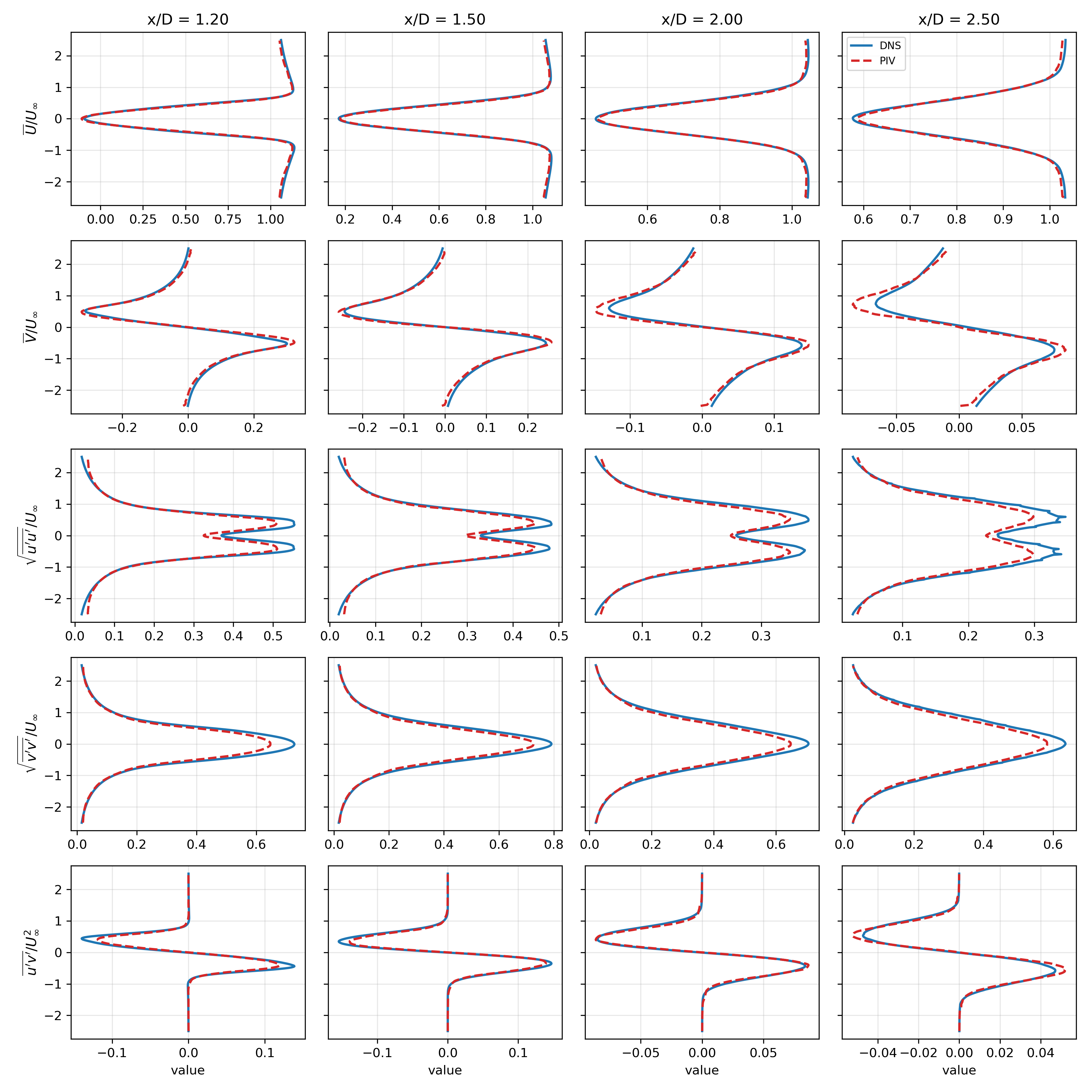}
  \caption{PIV (dashed) versus DNS (solid) for the circular cylinder at $\Rey=10^4$: wake
  profiles of $U/U_\infty$, $V/U_\infty$ and the in-plane Reynolds stresses (rows) at four wake
  stations $x/D=1.2,1.5,2.0,2.5$ (columns).}
  \label{fig:piv-circ}
\end{figure}

\begin{figure}
  \centering
  \includegraphics[width=0.85\textwidth]{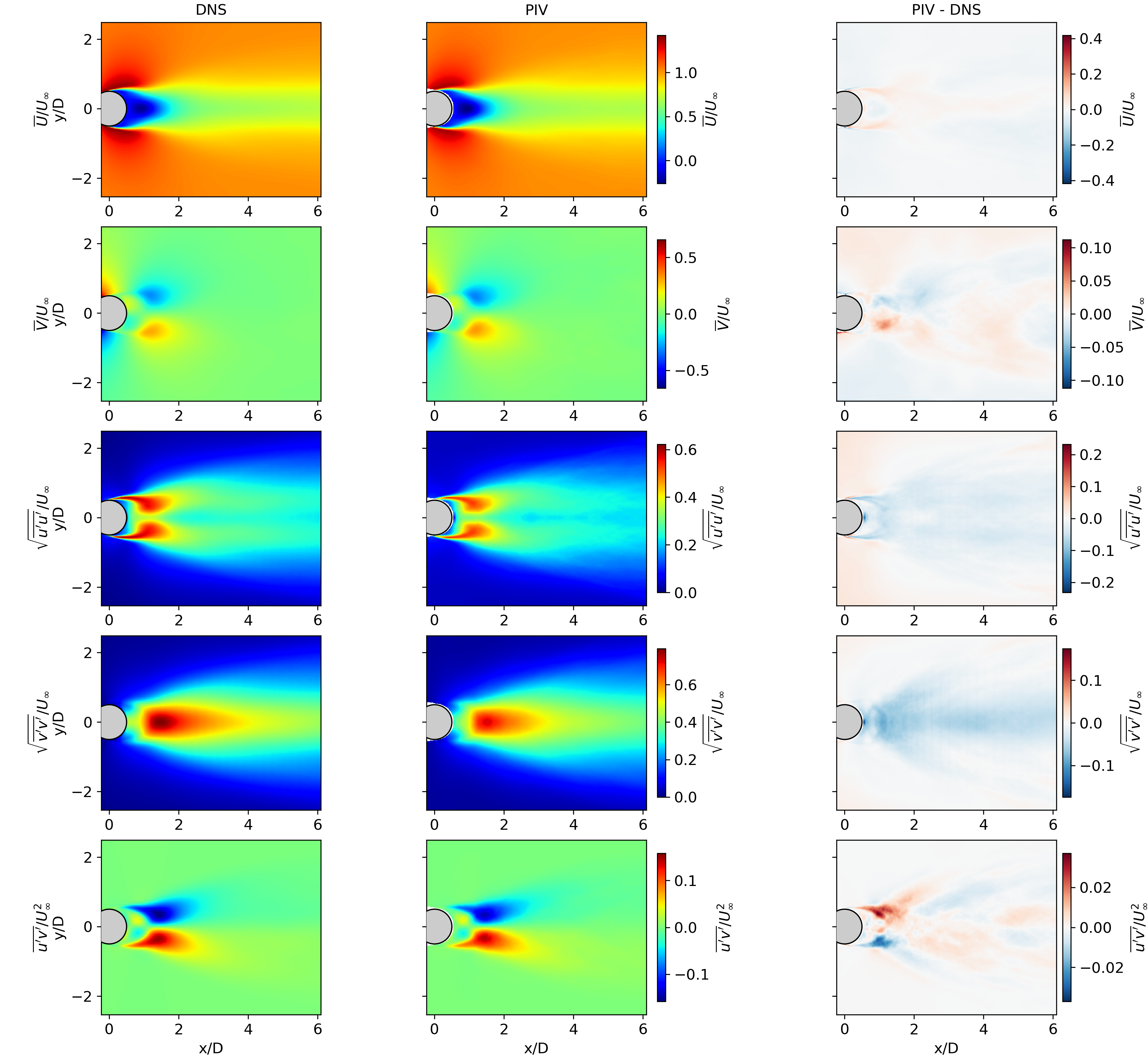}
  \caption{PIV versus DNS for the circular cylinder at $\Rey=10^4$ over the PIV measurement window.
  Columns: DNS, PIV, and their difference PIV$-$DNS (a near-zero difference column indicates
  agreement). Rows: the mean velocities $U/U_\infty$, $V/U_\infty$ and the in-plane Reynolds
  stresses $\sqrt{\langle u'^2\rangle}/U_\infty$, $\sqrt{\langle v'^2\rangle}/U_\infty$,
  $\langle u'v'\rangle/U_\infty^2$.}
  \label{fig:piv-circ-field}
\end{figure}

\rev{Planar PIV measurements provide an independent comparison for the circle,
diamond, and long triangle. The facility follows} \citet{zhang2026};
\rev{however, a high-resolution camera ( HZ-21000-G-C, Emergent Vision Technologies) was used, and the laser was positioned inside the tow tank, recording a separate continuous time series of data for each case; the acquisition and processing settings are listed in table~\ref{tab:dataset-setup}.
These measurements provide in-plane velocity statistics, but neither pressure nor
the out-of-plane velocity component.}
For the circular cylinder, the PIV agrees closely with DNS
both in profiles at successive wake stations (figure~\ref{fig:piv-circ}) and over the full
measurement window (figure~\ref{fig:piv-circ-field}); the corresponding profile and
field comparisons for the diamond and the long triangle are collected in appendix~\ref{app:piv}.

\rev{The conventional reference is the SST $k$--$\omega$ model}
\citep{menter1994}, \rev{solved in OpenFOAM for all six geometries. Its Reynolds stresses
are reconstructed through the Boussinesq relation.
The learned closures are deployed separately in a finite-element RANS solver
implemented in FEniCS, on body-fitted meshes refined around the body and in the wake, with symmetry imposed at $y=0$; their numerical
settings are described in the deployment methodology. Accuracy is assessed using
relative $L^2$ field errors against DNS and the drag coefficient $C_d$.}

\section{Methodology}\label{sec:method}

\rev{Section~\ref{sec:method-form} presents the governing equations and the stress and force
representations. Section~\ref{sec:method-pinn} defines the four closure models and their
solver-agnostic PINN training, and \S\ref{sec:method-solver} describes deployment in the
finite-element solver. Figure~\ref{fig:sketch} summarizes the overall pipeline.}

\subsection{\rev{Governing equations}}\label{sec:method-form}

For steady, incompressible, high-Reynolds-number flow we solve the Reynolds-averaged
Navier--Stokes (RANS) equations for the mean velocity $\bU=(U,V)$ and mean pressure $P$,
\rev{with $\bU$ measured relative to the body,}
\begin{equation}
  (\bU\!\cdot\!\nabla)\bU = -\nabla P + \nu\nabla^2\bU - \bnabla\!\cdot\tij,
  \qquad \nabla\!\cdot\bU = 0,
  \label{eq:rans}
\end{equation}
\rev{The unknown Reynolds stress $\tij=\langle u_i'u_j'\rangle$ is modelled either directly
(M1--M3) or through its divergence, the Reynolds force $\mathbf F=-\bnabla\!\cdot\tij$ (M4).
In both cases, neural networks determine the closure coefficients from the resolved mean field.}

\emph{Tensor-basis Reynolds stress.} In two dimensions the mean velocity gradient splits into the
symmetric strain-rate and antisymmetric rotation-rate tensors
\begin{equation}
  \Smat=\tfrac12(\nabla\bU+\nabla\bU^{\!\top})=\begin{pmatrix} a & b & 0\\ b & -a & 0 \\ 0 & 0 & 0\end{pmatrix},
  \qquad
  \Wmat=\tfrac12(\nabla\bU-\nabla\bU^{\!\top})=\begin{pmatrix} 0 & w & 0\\ -w & 0 & 0 \\ 0 & 0 & 0\end{pmatrix},
  \label{eq:SOmega}
\end{equation}
with $a=\tfrac12(U_x-V_y)$, $b=\tfrac12(U_y+V_x)$, $w=\tfrac12(U_y-V_x)$, and $\operatorname{tr}\Smat=0$
by incompressibility. Because the mean flow is two-dimensional but the turbulence is
three-dimensional, the stress is a $3\times3$ tensor ($\tau_{xz}=\tau_{yz}=0$ since $\partial_z=0$);
we represent it in the integrity basis of $(\Smat,\Wmat)$ augmented by an out-of-plane
redistribution tensor,
\begin{equation}
  \tij = \underbrace{\tfrac23 k\,\mathbf I}_{\text{isotropic}}
  \;\underbrace{-\,2\nut\Smat}_{\text{eddy viscosity}}
  \;+\;\underbrace{g_2\,\Tw}_{\text{in-plane anisotropy}}
  \;+\;\underbrace{\Gz\,\Tz}_{\text{out-of-plane}},
  \label{eq:tensorbasis}
\end{equation}
with basis tensors
\begin{equation}
  \rev{\Tw=\Smat\Wmat-\Wmat\Smat},
  \qquad
  \Tz=\mathrm{diag}\big(\tfrac16,\tfrac16,-\tfrac13\big),
  \label{eq:basis}
\end{equation}
and four scalar coefficients $(k,\nut,g_2,\Gz)$. Here $k\ge0$ is the turbulent kinetic energy
setting the trace, $\operatorname{tr}\tij=2k$; $\nut\ge0$ is the eddy viscosity; $g_2$ controls the
in-plane anisotropy; and $\Gz$ sets the in-plane/out-of-plane split. Both $\Tw$ and $\Tz$ are
traceless and energy-neutral---$\Tw{:}\Smat=0$ and $\Tz{:}\nabla\bU=\tfrac16\nabla\!\cdot\bU=0$---so
only the eddy-viscosity term does work on the mean flow.

\emph{The Reynolds force.} Equivalently, one may target the force the stress exerts on the mean
flow, $\mathbf F=-\bnabla\!\cdot\tij=(F_x,F_y)$, which closes momentum directly,
\begin{equation}
  (\bU\!\cdot\!\nabla)\bU = -\nabla P + \nu\nabla^2\bU + \mathbf F,
  \qquad \nabla\!\cdot\bU = 0.
  \label{eq:rans-force}
\end{equation}
The force does work $\bU\!\cdot\!\mathbf F$ on the mean flow, so an admissible force closure should
respect that energy budget. \rev{We use an eddy-viscosity contribution and a regularized residual force,}
\begin{equation}
  \mathbf F=\underbrace{\bnabla\!\cdot(2\nut\Smat)}_{\text{dissipative}}
  \;+\;\underbrace{\mathbf R_\perp}_{\rev{\text{regularized residual}}},
  \label{eq:forcesplit}
\end{equation}
\rev{where the residual is exactly energy-neutral only in the unregularized limit
at nonzero velocity; the implemented projection is given in \eqref{eq:perp}.}

\subsection{\rev{Solver-agnostic PINN training of closure models}}\label{sec:method-pinn}

\begin{figure}
  \centering
  \includegraphics[width=0.7\textwidth]{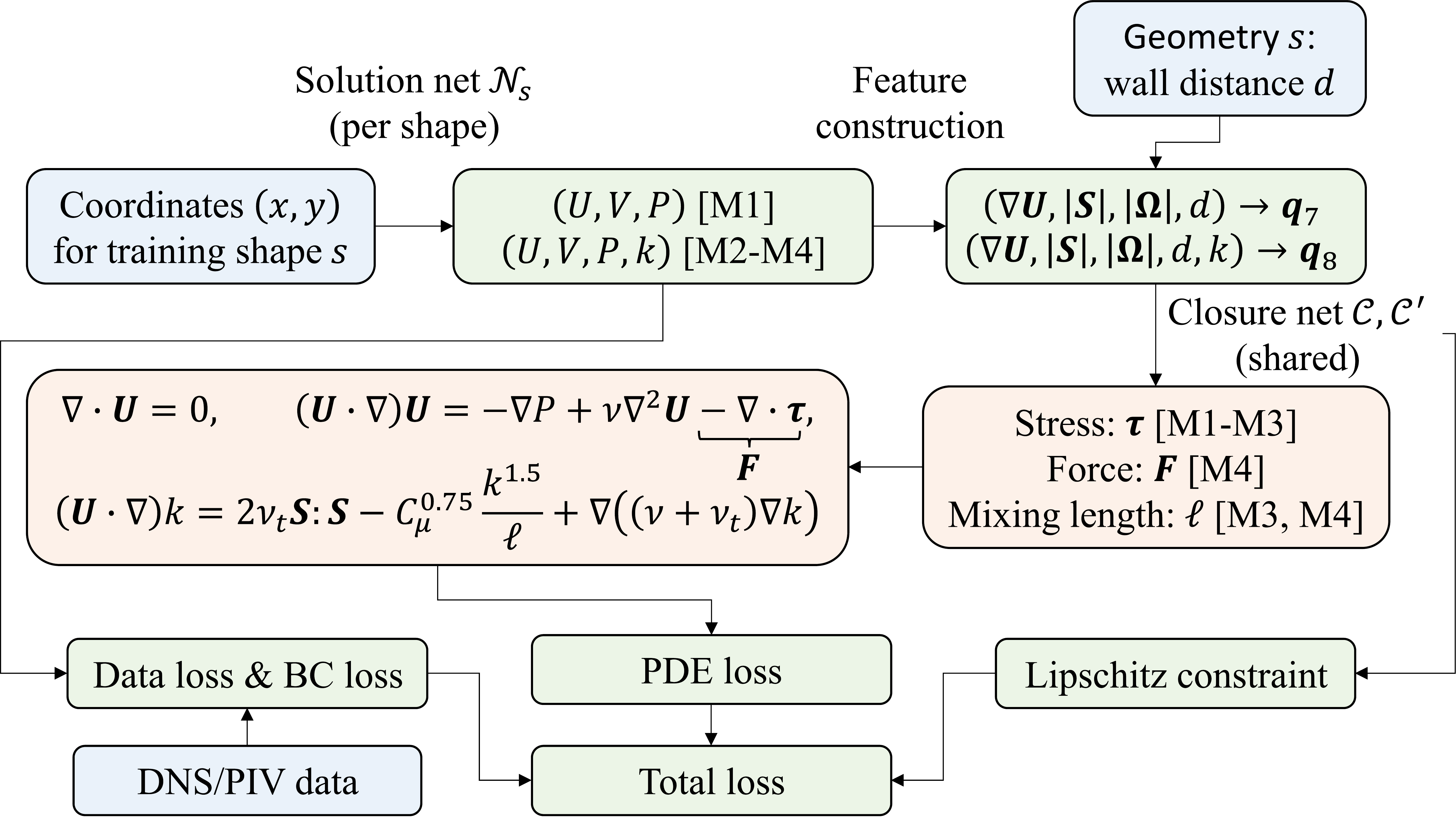}
  \caption{\rev{PDE-constrained identification of neural turbulence closures.
  Each training geometry has a separate solution network $\mathcal N_s$, while the closure
  networks $\mathcal C$ and $\mathcal C'$ are shared
  across geometries. 
  Table~\ref{tab:features} details the input features and their normalization;
  table~\ref{tab:nets} gives the network mappings and closure forms for M1--M4.}}
  \label{fig:pinn-closure-training}
\end{figure}

\rev{Figure~\ref{fig:pinn-closure-training} shows the diagram of PDE-constrained PINN training of closures. Each geometry $s$ has a solution network $\mathcal N_s$ predicting $(U,V,P)$
for M1 and $(U,V,P,k)$ for M2--M4. The closure network $\mathcal C$ and, for M3/M4,
the length-scale network $\mathcal C'$ are shared across training geometries and optimized
jointly with the solution networks. Only the shared networks are retained for deployment.}

\rev{The closure inputs are the seven kinematic features}
\begin{equation}
  \mathbf q_7 = \big(U_x,\,U_y,\,V_x,\,V_y,\,|\Smat|,\,|\Wmat|,\,d\big)\in\mathbb R^7,
  \label{eq:features}
\end{equation}
\rev{with transported $k$ appended for M2--M4 to give $\mathbf q_8=(\mathbf q_7,k)$.
Table~\ref{tab:features} specifies their normalization, and table~\ref{tab:nets}
summarizes the four closure mappings.}

\begin{table}
  \centering
  \caption{Input features of the closure networks. Each feature enters through the bounded limiter
  $\hat f=f/(|f|+f_{\mathrm{ref}})\in(-1,1)$ (wall distance through $\tanh(d/d_{\mathrm{ref}})$), with
  $f_{\mathrm{ref}}$ the natural unit ($=1$ under $U_\infty=D=1$). The seven kinematic features form
  $\mathbf q_7$; M2--M4 append the transported $k$ to form $\mathbf q_8=(\mathbf q_7,k)$.}
  \label{tab:features}
  \begin{tabular}{lllc}
    \toprule
    feature & symbol & normalized input $\hat f$ & $f_{\mathrm{ref}}$ \\
    \midrule
    \multicolumn{4}{l}{\textit{Kinematic features $\mathbf q_7$ (all methods)}}\\
    velocity-gradient components & $U_x,U_y,V_x,V_y$ & $f/(|f|+S_{\mathrm{ref}})$ & $S_{\mathrm{ref}}=U_\infty/D$ \\
    strain-rate magnitude        & $|\Smat|$ & $|\Smat|/(|\Smat|+S_{\mathrm{ref}})$ & $S_{\mathrm{ref}}=U_\infty/D$ \\
    rotation-rate magnitude      & $|\Wmat|$ & $|\Wmat|/(|\Wmat|+S_{\mathrm{ref}})$ & $S_{\mathrm{ref}}=U_\infty/D$ \\
    signed wall distance         & $d$ & $\tanh(d/d_{\mathrm{ref}})$ & $d_{\mathrm{ref}}=D$ \\
    \midrule
    \multicolumn{4}{l}{\textit{Non-local feature appended for $\mathbf q_8$ (M2--M4)}}\\
    transported TKE              & $k$ & $k/(k+k_{\mathrm{ref}})$ & $k_{\mathrm{ref}}=U_\infty^2$ \\
    \bottomrule
  \end{tabular}
\end{table}

\begin{table}
  \centering
  \caption{\rev{Network mappings and closure reconstruction. The closure chains show
  normalized inputs, neural-network outputs, and reconstructed stress or force.
  M2--M4 also use transported $k$ and the indicated length scale.}}
  \label{tab:nets}
  \footnotesize
  \setlength{\tabcolsep}{3pt}
  \begin{tabular}{llll}
    \toprule
    method & solution net $\mathcal N_s$ & \rev{closure chain} & \rev{length scale} \\
    \midrule
    M1 & $(x,y)\mapsto(U,V,P)$ & \rev{$\mathbf q_7\xrightarrow{\mathcal C}(k,\nut,g_2)\longrightarrow\tij$} & --- \\
    M2 & $(x,y)\mapsto(U,V,P,k)$ & \rev{$\mathbf q_8\xrightarrow{\mathcal C}(m,g_2,\Gz)\longrightarrow\tij$} & \rev{$\lmix=\min(\kvk d,\lmix^{\max})$} \\
    M3 & $(x,y)\mapsto(U,V,P,k)$ & \rev{$\mathbf q_8\xrightarrow{\mathcal C}(m,g_2,\Gz)\longrightarrow\tij$} & \rev{$\mathbf q_7\xrightarrow{\mathcal C'}z_\ell\to\lmix$} \\
    M4 & $(x,y)\mapsto(U,V,P,k)$ & \rev{$\mathbf q_8\xrightarrow{\mathcal C}(m,R_x,R_y)\longrightarrow\mathbf F$} & \rev{$\mathbf q_7\xrightarrow{\mathcal C'}z_\ell\to\lmix$} \\
    \bottomrule
  \end{tabular}
\end{table}

\medskip\noindent\rev{\emph{Closure forms.} The following mappings define M1--M4;
realizability constraints, output transformations, and parameter ranges are given in
Appendix~\ref{app:closure-constraints}.}

\rev{\emph{M1: local stress.} The closure network $\mathcal C$ maps $\mathbf q_7$ to $(k,\nut,g_2)$,
where $k=\tfrac12\langle u'^2+v'^2\rangle$ is the in-plane turbulent kinetic energy. Its in-plane stress is reconstructed as}
\begin{equation}
  \rev{\tij_{\parallel}=k\mathbf I_2-2\nut\Smat_{\parallel}+g_2(\Tw)_{\parallel},}
  \label{eq:M1-reconstruction}
\end{equation}
\rev{where $\mathbf I_2$ is the $2\times2$ identity matrix and the subscript $\parallel$ denotes the in-plane block.}

\rev{\emph{M2: transported-$k$ stress.} The closure network $\mathcal C$ maps
$\mathbf q_8$ to $(m,g_2,\Gz)$. The solution network supplies the full turbulent kinetic
energy $k=\tfrac12\langle u'^2+v'^2+w'^2\rangle$, constrained by}
\begin{equation}
  \rev{(\bU\!\cdot\!\nabla)k
  =2\nut(\Smat{:}\Smat)-\Cmu^{3/4}\frac{k^{3/2}}{\lmix}
  +\nabla\!\cdot\!\big[(\nu+\nut)\nabla k\big].}
  \label{eq:keq}
\end{equation}
\rev{An algebraic mixing length determines the eddy viscosity,}
\begin{equation}
  \nut=\Cmu^{1/4}m\sqrt{k+\varepsilon}\,\lmix,
  \qquad \lmix=\min(\kvk d,\lmix^{\max}),
  \label{eq:nut}
\end{equation}
\rev{and the stress is reconstructed as
$\tij=\tfrac23 k\mathbf I-2\nut\Smat+g_2\Tw+\Gz\Tz$, as in \eqref{eq:tensorbasis}.}

\rev{\emph{M3: learned-length-scale stress.} The closure network $\mathcal C$ maps
$\mathbf q_8$ to $(m,g_2,\Gz)$, with the same stress reconstruction as M2.
An auxiliary closure network $\mathcal C'$ maps $\mathbf q_7$ to a scalar $z_\ell$,
from which the mixing length is constructed:}
\begin{equation}
  \rev{z_\ell=\mathcal C'(\mathbf q_7),\qquad \lmix=\mathcal H(d,z_\ell),}
  \label{eq:length-map}
\end{equation}
\rev{where the bounded transformation $\mathcal H$ is defined in \eqref{eq:learnedell}.
This length enters both the destruction term of the $k$-transport equation \eqref{eq:keq} and the eddy viscosity $\nut$ in \eqref{eq:nut}. The auxiliary network does not receive $k$.}

\rev{\emph{M4: Reynolds force.} The closure network $\mathcal C$ maps $\mathbf q_8$ to $(m,R_x,R_y)$.
The transported $k$, eddy viscosity, and learned length scale follow M3. With
$\mathbf R=(R_x,R_y)$, the force is reconstructed through the regularized projection}
\begin{equation}
  \rev{\mathbf R_\perp=\mathbf R-\frac{\mathbf R\!\cdot\!\bU}{|\bU|^2+\varepsilon_p}\bU,
  \qquad \mathbf F=\bnabla\!\cdot(2\nut\Smat)+\mathbf R_\perp.}
  \label{eq:perp}
\end{equation}
\rev{Here $\varepsilon_p=10^{-3}$ in nondimensional units. The projection remains
finite at $\bU=0$, but satisfies
$\mathbf R_\perp\cdot\bU=\varepsilon_p(\mathbf R\cdot\bU)/(|\bU|^2+\varepsilon_p)$,
so it does not enforce exact orthogonality at finite regularization.
The constraint concerns work relative to the body: $\bU=\bU_{\rm lab}-\bU_{\rm body}$
is unchanged by a uniform Galilean boost of both velocities.}

\rev{\emph{Training objective.} The total loss is summed over training geometries,}
\begin{equation}
  \rev{\mathcal L=\sum_{s\in\mathcal S_{\mathrm{train}}}\mathcal L_s+\lambda_{\mathrm{lip}}\mathcal L_{\mathrm{lip}},}
  \label{eq:loss}
\end{equation}
\rev{with the objective for each geometry given by}
\begin{equation}
  \rev{\mathcal L_s=\mathcal L_{\mathrm{pde},s}+\mathcal L_{\mathrm{data},s}
  +\mathcal L_{\mathrm{bc},s}.}
  \label{eq:Lgroups}
\end{equation}
\rev{The PDE, data, and boundary terms are evaluated separately for each geometry;
the Lipschitz term regularizes the shared closure across all training shapes.
The sum in \eqref{eq:Lgroups} is not unweighted: each group carries its own adaptive
weights, $w_{r,s}$ and $w_{rk,s}$ in \eqref{eq:Lpde}, and $w_U$, $w_k$, $w_\tau$, $w_F$,
and $w_b$ in the data and boundary losses (Appendix~\ref{app:loss-weighting}).}

\rev{\emph{PDE loss.} The spatial residuals for each geometry are}
\begin{equation}
\rev{\begin{aligned}
  r_c &= \nabla\!\cdot\bU, &&\text{M1--M4},\\
  \mathbf r_m &= (\bU\!\cdot\!\nabla)\bU+\nabla P-\nu\nabla^2\bU
  +\begin{cases}\nabla\!\cdot\tij,&\text{M1--M3},\\-\mathbf F,&\text{M4},\end{cases}\\
  r_k &= (\bU\!\cdot\!\nabla)k-2\nut(\Smat{:}\Smat)
  +\Cmu^{3/4}\frac{k^{3/2}}{\lmix}
  -\nabla\!\cdot\!\big[(\nu+\nut)\nabla k\big], &&\text{M2--M4}.
\end{aligned}}
\label{eq:residuals}
\end{equation}
\rev{Writing $\mathbf r_m=(r_u,r_v)$, the PDE loss is}
\begin{equation}
  \rev{\mathcal L_{\mathrm{pde},s}
  =w_{r,s}\big\langle r_c^2+\tilde r_u^2+\tilde r_v^2\big\rangle_s
  +\mathbf 1_{\mathrm{M2--M4}}w_{rk,s}\big\langle\tilde r_k^2\big\rangle_s.}
  \label{eq:Lpde}
\end{equation}
\rev{Here $\langle\cdot\rangle_s$ averages over collocation points in geometry $s$,
and $\mathbf 1_{\mathrm{M2--M4}}$ excludes the $k$ residual for M1.
Spatial derivatives are obtained by automatic differentiation. The tildes denote
pseudo-time relaxation (Appendix~\ref{app:training-residuals}); continuity is imposed directly.}

\rev{\emph{Data loss.} All models fit the observed velocity and kinetic energy,
using in-plane $k$ for M1 and full $k$ for M2--M4. M1 fits
$(\tau_{xx},\tau_{xy},\tau_{yy})$; M2/M3 fit all nonzero independent stress components, $(\tau_{xx},\tau_{xy},\tau_{yy},\tau_{zz})$.
M4 replaces stress supervision with force supervision when force data are available.
Only observations available in the selected DNS or PIV training protocol enter the loss.
The data loss is a weighted mean-squared misfit against the targets $(\cdot)_\star$,}
\begin{equation}
  \rev{\mathcal L_{\mathrm{data},s}
  = w_{U}\big\langle\|\bU-\bU_\star\|^2\big\rangle_s
  + w_{k}\big\langle(k-k_\star)^2\big\rangle_s
  + w_{\tau}\Big\langle\textstyle\sum_{ij}(\tau_{ij}-\tau_{ij\star})^2\Big\rangle_{s}
  + w_{F}\big\langle\|\mathbf F-\mathbf F_\star\|^2\big\rangle_{s},}
  \label{eq:Ldata}
\end{equation}
\rev{where the stress term is present for M1--M3, summed over the fitted components, and
the force term for M4. The DNS force targets $\mathbf F_\star=-\nabla\!\cdot\boldsymbol\tau_\star$
are obtained by finite differences of the DNS Reynolds stresses on a uniform grid,
Gaussian-filtered, and interpolated back to the observation points.}

\rev{\emph{Boundary loss.} All models impose uniform inflow, zero-pressure outflow,
far-field symmetry, and no-slip on the body. M2--M4 also impose $k=0$ on the body and inlet.}

\rev{\emph{Lipschitz loss.} The shared regularizer is}
\begin{equation}
  \rev{\mathcal L_{\mathrm{lip}}
  =\left\langle\left\|\frac{\partial\mathbf c}{\partial\mathbf q_7}\right\|_F^2\right\rangle_{\mathrm{train}},
  \qquad \lambda_{\mathrm{lip}}=0.03,}
  \label{eq:Llip}
\end{equation}
\rev{where $\mathbf c$ denotes the closure coefficient outputs for M1--M3 and the
residual force $\mathbf R$ for M4. The average uses feature samples from all training shapes;
$k$ is held fixed when differentiating the non-local closures with respect to $\mathbf q_7$.
The PDE, data, and boundary weights are adapted by gradient-norm balancing
(Appendix~\ref{app:loss-weighting}).}

\rev{Table~\ref{tab:training-configuration} in Appendix~\ref{app:training-setup} summarizes the networks, optimizer,
sampling, and computational cost. Smooth $\tanh$ activations support the spatial
derivatives required by the PDE residuals.}

\emph{Cost versus a solver-in-the-loop baseline.} The absence of a forward solve in the training
loop is what makes this fast. For comparison, obtaining a data-driven Reynolds-force model of
comparable scope by the conventional adjoint route takes of order a day: the
OpenFOAM-based adjoint optimization we used previously \citep{zhang2026} required $\sim$40 adjoint
iterations and $\sim$200 forward RANS (SIMPLE) evaluations on 8 CPU cores, the wall-clock split
between two costs per iteration---the forward solves demanded by the optimizer's line search, and the
GMRES solution of the saddle-point adjoint system---and this counts only the optimization run, not the
one-off effort of deriving and implementing the adjoint for that specific force model. \rev{Taking one day as a rough baseline, the 15--67 min PINN training times
correspond to an indicative wall-clock speedup of approximately $20$--$100\times$
(about $100\times$ for single-shape training and $20\times$ for five-shape training).
This comparison uses the reported CPU and GPU configurations and different training scopes;
it is not a matched-hardware benchmark.}

\subsection{Finite-element deployment and stabilization}\label{sec:method-solver}

We deploy the frozen $\mathcal{C}$ in an independent finite-element RANS solver implemented in FEniCS
\citep{alnaes2015fenics}---a Taylor--Hood ($P_2/P_1$) discretization \citep{taylor1973} of
\eqref{eq:rans} on a \rev{body-fitted triangular mesh}, solved by
Picard iteration on the half-domain ($y\ge0$) with a symmetry condition at the wake centreline. For
the stress closures the eddy-viscosity part of \eqref{eq:tensorbasis} is treated implicitly (as
turbulent diffusion) and the remaining $\tfrac23 k\mathbf I + g_2\Tw + \Gz\Tz$ as an explicit source,
in weak-divergence form so the network is never strong-differentiated; at convergence the stress
equals the modelled $\tij$ while the iteration retains the stabilizing implicit diffusion. For M4 the
eddy-viscosity diffusion $\bnabla\!\cdot(2\nut\Smat)$ is likewise treated implicitly, while the
\rev{regularized residual $\mathbf R_\perp$ is added} as an explicit body force, so the converged momentum
solves the force balance \eqref{eq:rans-force} with $\nut$ supplying the stabilizing implicit
diffusion. For the
non-local variants the $k$-equation \eqref{eq:keq} is solved as a coupled transport equation in the
same fixed-point loop: each Picard step updates the closure from $\nabla\bU$ and the current $k$,
solves momentum and continuity, then solves a stabilized weak form of \eqref{eq:keq}
(destruction linearized as $\Cmu^{3/4}k_n^{1/2}k/\lmix$, with $k=0$ on the body and inlet),
under-relaxing $k$ like $\bU$.

Naive coupling of a frozen data-driven stress is ill-conditioned \citep{wu2019}: small stress
errors can drive large velocity errors, and the Picard iteration may stall or diverge. We stabilize
the deployed solve with three devices, ablated in \S\ref{sec:results-stab}: (i)~input-gradient
smoothing---\rev{a Helmholtz filter with a physical, strain-dependent length applied
before the velocity gradients enter the closure}; (ii)~a standard streamline-upwind (SUPG) treatment \citep{brooks1982supg} of the
convection operator; and (iii)~the closure-Lipschitz penalty ($\lambda_{\mathrm{lip}}=0.03$) imposed
during training. The same stabilized configuration is used for all four closures (M1--M4). \rev{For each raw gradient component $g\in\{U_x,U_y,V_x,V_y\}$, the filtered input
$\bar g$ is defined by}
\begin{equation}
  \rev{\bar g-\nabla\!\cdot(\delta_f^2\nabla\bar g)=g,
  \qquad \frac{\delta_f}{D}=\frac{C_f}{1+sD/U_\infty},
  \qquad C_f=0.2,}
  \label{eq:physical-filter}
\end{equation}
\rev{with zero normal flux on the boundary. Here
$s=\sqrt{2\mathbf S_f{:}\mathbf S_f}$ and
$\mathbf S_f=\tfrac12(\nabla\bU+\nabla\bU^{\mathsf T})$ are evaluated from the current
unfiltered velocity and updated at each Picard iteration; $\delta_f$ is independent of
the cell size.
The idea is to present the closure with inputs of the same smoothness as those it was trained
on. During training, the closure sees gradients of the smooth network fields, whereas the
discrete velocity gradients in the solver are piecewise polynomial, discontinuous across
elements and contaminated by grid-scale oscillations, particularly in the shear layers and near
separation. Because the closure is nonlinear and was never trained on such inputs, it amplifies
these oscillations into the stress source; the source then perturbs the velocity at the next
Picard iteration, and the loop stalls instead of contracting. The Helmholtz operator in
\eqref{eq:physical-filter} acts as a low-pass filter: a Fourier mode of wavenumber $\kappa$ is
attenuated by the factor $1/(1+\delta_f^2\kappa^2)$, so variations on scales below $\delta_f$
are damped, while larger scales pass almost unchanged. The width adapts to the local strain. In
weakly strained regions, such as the free stream and the far wake, $\delta_f\to C_fD$ and
residual oscillations are removed. In strongly strained regions ($sD/U_\infty\gg1$),
$\delta_f\approx C_fU_\infty/s$, a fixed fraction of the local strain length $U_\infty/s$, so
the thin boundary and shear layers that carry the physical gradients are not smeared. Since
$\delta_f$ is defined by the flow rather than by the mesh, the filtered input converges to a
well-defined continuum quantity under refinement. Only the closure input is filtered; the
momentum and continuity equations use the unfiltered velocity, so the discrete conservation
form is unaffected.
For the in-sample M1 cylinder (Appendix~\ref{app:filter-sensitivity}), varying
$C_f$ over $[0.05,0.4]$ changes the velocity by at most $2.3\%$ relative to $C_f=0.2$, the
pressure by up to $12\%$ and the stresses by up to $18\%$, while $C_d$ increases from
$1.10$ to $1.35$ (DNS: $1.21$) and the iteration count falls from 236 to 83. $C_f$ is thus a
modelling parameter of the deployed closure; we fix $C_f=0.2$ for all closures and meshes.}

\rev{\emph{Nonlinear convergence.} Convergence is monitored with the mass-weighted
residual of each equation---the two momentum components, continuity and, when present, the
$k$ equation---normalized by its initial value. The convergence threshold is the relative
residual of $10^{-5}$.}

\rev{We use velocity under-relaxation of $0.4$, an eddy-viscosity cap $\nut\le0.5$,
and the medium-mesh spacings
$(h_{\mathrm{wall}},h_{\mathrm{wake}},h_{\mathrm{far}})/D=(0.02,0.08,0.8)$.
For the in-sample M1 cylinder, doubling and halving these spacings (2690, 9708 and 37059
half-domain cells) changes the velocity by $0.78\%$ (coarse--medium) and $0.33\%$
(medium--fine), and the pressure by $3.9\%$ and $1.4\%$, so the medium mesh is used throughout.}

\section{Results}\label{sec:results}

All four closures of \rev{\S\ref{sec:method-pinn}}---the three tensor-basis stress closures M1 (local),
M2 (non-local transported-$k$ with algebraic length scale) and M3 (non-local with a learned $k$-free
length scale), and the forcing-vector model M4---are trained on equal footing inside the PINN of
\S\ref{sec:method-pinn} and then evaluated by deployment in the independent finite-element solver of
\S\ref{sec:method-solver}.
The section follows the chain from training to deployment. We first establish why a
PINN-trained closure deploys at all: imposing the RANS residual makes it PDE-consistent rather
than a mere a-priori fit, so its accuracy survives transfer to an independent solver
(\S\ref{sec:results-apriori}). We then isolate the numerical stabilization a frozen data-driven
closure needs in that solver, finding input-gradient smoothing decisive (\S\ref{sec:results-stab}),
and show that the same solver-agnostic trainer is so data-light that it learns directly from patched
experimental PIV in place of DNS (\S\ref{sec:results-piv}). With deployment secured, we report
quantitative accuracy in two regimes---in-sample, each closure trained on its own shape
(\S\ref{sec:results-insample}), and a strict leave-one-shape-out test of geometry
generalization (\S\ref{sec:results-loso})---finding M3 the most accurate on the stress field and the
force model M4 the best generalizer on the mean flow and drag.
\rev{Throughout, $\mathcal E(q)=\|q-q^{\mathrm{DNS}}\|_{2}/\|q^{\mathrm{DNS}}\|_{2}$
denotes the relative $L^2$ error of a field $q$ against DNS (or PIV in
\S\ref{sec:results-piv}), evaluated over the comparison window with pressure
gauge-shifted to zero mean.}

\subsection{A priori versus a posteriori: a PDE-consistent closure}\label{sec:results-apriori}
A learned closure is only useful once it is substituted back into a solver, and how training secures
that property is the key distinction here. Classical data-driven methods make the closure
solver-consistent: they place a specific CFD solver in the optimization loop so the model is
calibrated directly against that solver's output---effective, but tying the
closure to one discretization. Pure a-priori fitting takes the opposite tack, regressing the
closure directly against the stored Reynolds stress, optimizing agreement with the data rather
than with the governing equations; such a fit need not deploy. Our PINN occupies neither extreme: by
imposing the RANS residual in training it makes the closure PDE-consistent---consistent with
the governing equations themselves rather than with any one solver---which is exactly the property
that lets it transfer to an independent solver. The closure is therefore not a simple a-priori fit;
PDE-consistency is a solver-diagnostic guarantee, the conceptual core of the method.

\begin{figure}
  \centering
  \includegraphics[width=\textwidth]{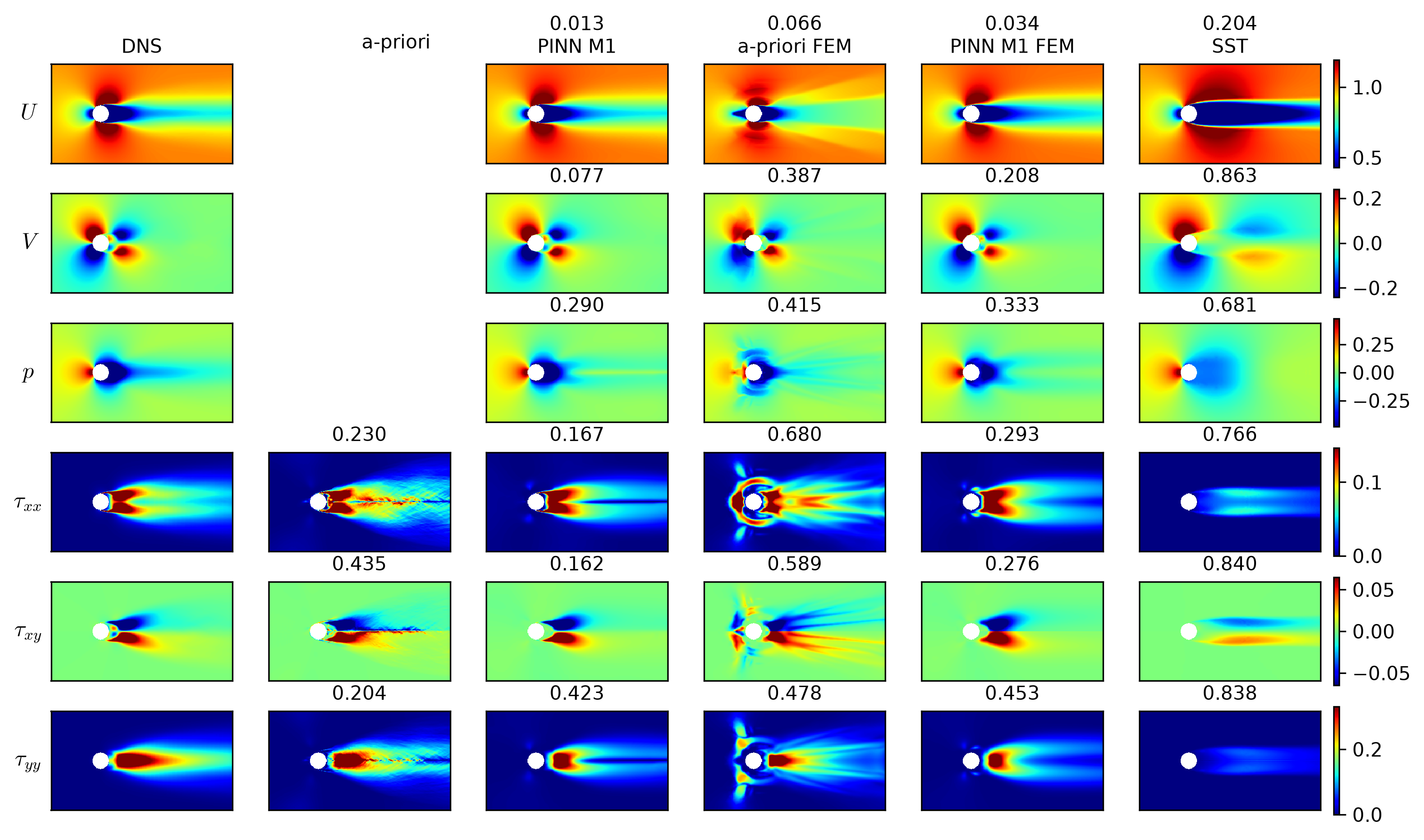}
  \caption{A priori fittting versus PINN training for M1 (in-sample cylinder). Columns: DNS; the
  a-priori fit (pure stress regression, no mean field); the supervised-PINN M1; their
  two FEM deployments; SST $k$--$\omega$. Rows $U,V,p,\tau_{xx},\tau_{xy},\tau_{yy}$; \rev{the number
  above each panel is the full-field relative $L^2$ error $\mathcal E(q)$ of that field against DNS.}}
  \label{fig:exp-apriori}
\end{figure}

Figure~\ref{fig:exp-apriori} makes this concrete on the local model M1, contrasting two trainings of
the same closure on the in-sample cylinder: a pure-supervised regression against the stored
DNS stress with no RANS residual (the classical a-priori fit), and our supervised-PINN
training, which adds the RANS residual to the same data supervision. The a-priori fit has no mean
field of its own---its $U,V,p$ panels are blank, only $\tau$ is defined---and although its read-out
stress is reasonable (\rev{$\mathcal E(\tau_{xx})=0.230$}), once it is frozen and deployed in the FEM solver its
accuracy does not survive: the stress fills with spurious cell-scale striations and collapses
(\rev{$\mathcal E(\tau_{xx})=0.680$}, \rev{$\mathcal E(\tau_{xy})=0.589$}), dragging the mean field with it (\rev{$\mathcal E(U)=0.066$}, \rev{$\mathcal E(V)=0.387$},
\rev{$\mathcal E(p)=0.415$}). The supervised-PINN closure behaves oppositely: its a-priori network field (\rev{$\mathcal E(U)=0.013$},
\rev{$\mathcal E(\tau_{xx})=0.167$}) and its a-posteriori deployment (\rev{$\mathcal E(U)=0.034$}, \rev{$\mathcal E(\tau_{xx})=0.293$}) stay close---the
small a-priori-to-a-posteriori gap that is the signature of PDE-consistency \citep{duraisamy2021}---and
once deployed it tracks DNS better than both \rev{the deployed a-priori fit} and the SST baseline on every field
(mean velocity \rev{$\mathcal E(U)=0.034$}, against $0.066$ for the deployed a-priori fit and $0.204$ for SST).
The figure is thus direct evidence that imposing the RANS residual in training, not the data fit
alone, is what lets the closure transfer to an independent solver. The supervised-PINN a-priori
fields for all four closures, and an \rev{inverse-training ablation for M3}, are collected in
appendix~\ref{app:pinn}.

\subsection{Solver stabilization in the a-posteriori deployment}\label{sec:results-stab}
\begin{figure}
  \centering
  \includegraphics[width=0.85\textwidth]{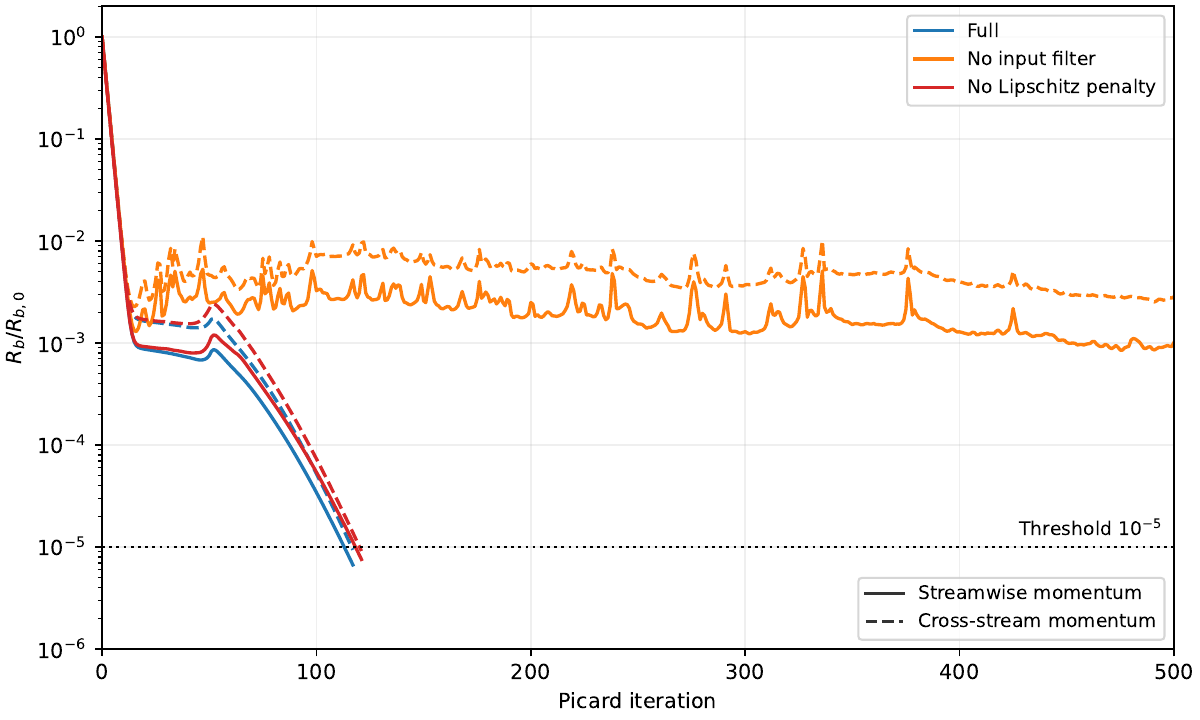}
  \caption{\rev{Relative momentum residuals of the M1 deployment on the in-sample cylinder
  (solid: streamwise; dashed: cross-stream) for the full configuration, no input filter,
  and no Lipschitz penalty. The dotted line marks the threshold $10^{-5}$.}}
  \label{fig:exp-picard}
\end{figure}

\rev{Figure~\ref{fig:exp-picard} shows the complete momentum-residual histories for
the Picard deployment of M1 on the in-sample cylinder, using the medium mesh and
convergence criteria of \S\ref{sec:method-solver}. Each equation residual is normalized
by its initial value. The full configuration, with the physical input filter
($C_f=0.2$) and Lipschitz regularization, satisfies the relative $10^{-5}$ threshold
after 117 iterations. Without input filtering,
the solve does not converge within 500 iterations: the final normalized streamwise
and cross-stream momentum residuals are $1.01\times10^{-3}$ and $2.84\times10^{-3}$.
Removing the Lipschitz penalty increases the iteration count modestly, from 117 to 121.
Removing SUPG also gives convergence in 117 iterations with nearly identical histories,
so that curve is omitted. Thus the physical input filter is necessary for convergence
in this test, whereas Lipschitz regularization has a smaller effect on the iteration count.}

\subsection{Training from patched PIV data}\label{sec:results-piv}
\begin{figure}
  \centering
  \includegraphics[width=0.86\textwidth]{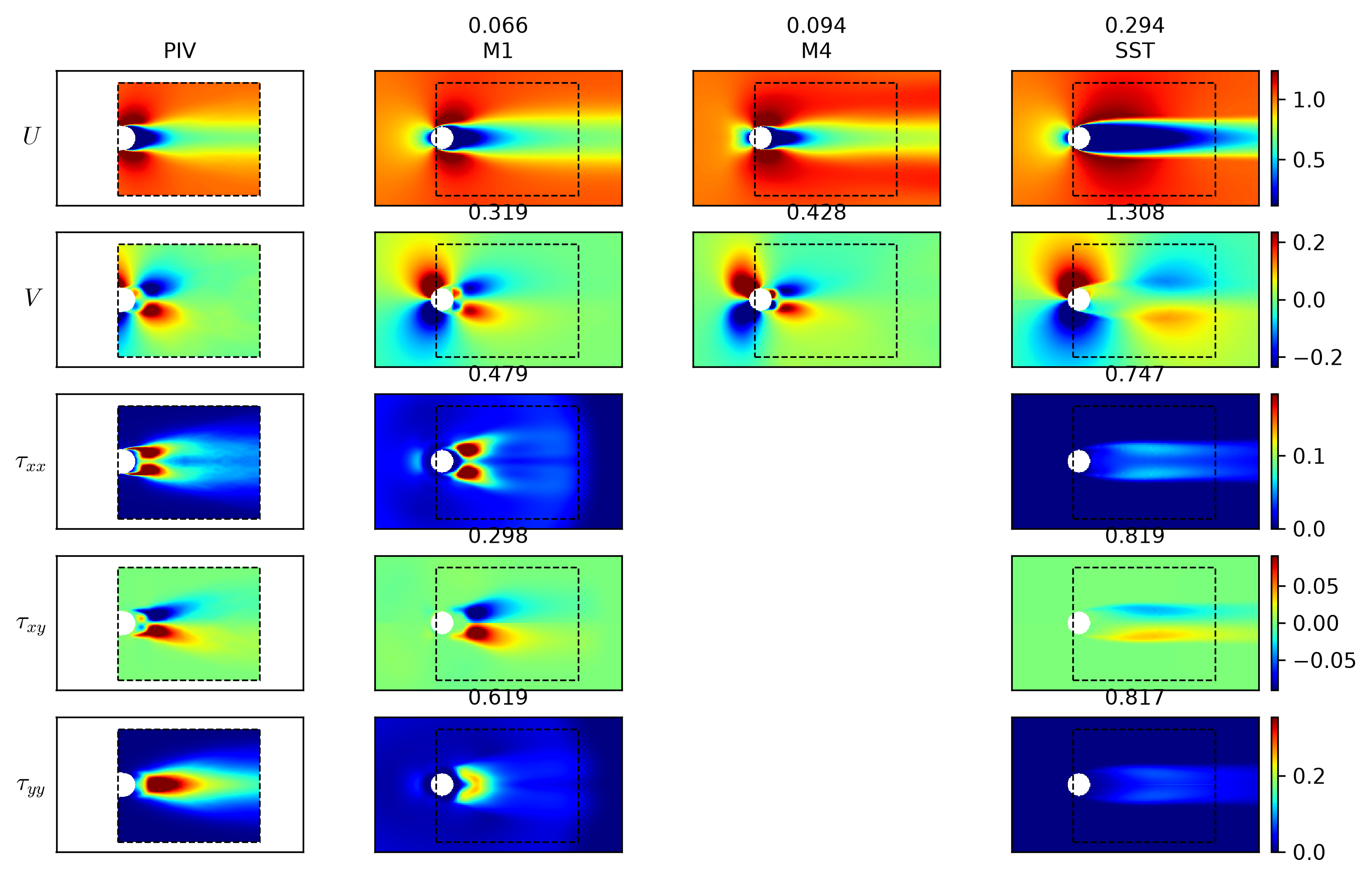}
  \caption{PIV-trained closures on the cylinder, FEM-deployed and compared to the PIV field over its
  validity window (dashed box). Columns PIV, M1, M4, SST; rows $U,V,\tau_{xx},\tau_{xy},\tau_{yy}$
  (M4 stores no stress); panel titles give rel-$L^2$ vs PIV.}
  \label{fig:exp-piv}
\end{figure}

Figure~\ref{fig:exp-piv} shows that the DNS supervision used above can be replaced by patched,
experiment-like measurements: because the PINN imposes the RANS residual everywhere and never runs a
solver, it needs only enough data to anchor the mean field, not a fully resolved DNS. The PINN is
otherwise unchanged---the same large computational domain and boundary conditions of
\S\ref{sec:method-pinn}, with only the data supervision restricted to the patched PIV window.
Particle-image velocimetry (PIV) is an in-plane, two-dimensional measurement (no pressure, no
out-of-plane $\langle w'^2\rangle$), so we train the local model M1 and the force model
M4---not M2 or M3, which directly supervise the out-of-plane stress $\tzz$ that PIV cannot provide---on the
cylinder PIV, FEM-deploy them, and compare to the measured field over its validity window (the dashed
box). The in-plane local closure M1 matches the PIV best, recovering the streamwise velocity to
\rev{$\mathcal E(U)=0.066$} and the cross-stream component to \rev{$\mathcal E(V)=0.319$}, ahead of the force model M4 (\rev{$\mathcal E(U)=0.094$},
\rev{$\mathcal E(V)=0.428$}) and well ahead of SST (\rev{$\mathcal E(U)=0.294$}, $E(V)=1.308$); it also reproduces the in-plane Reynolds
stresses (\rev{$\mathcal E(\tau_{xx})=0.479$}, \rev{$\mathcal E(\tau_{xy})=0.298$}, \rev{$\mathcal E(\tau_{yy})=0.619$}), each well below the corresponding
SST errors ($0.747$, $0.819$, $0.817$), whereas M4 stores no stress. The gap between M4 and M1 is
partly a data limitation: M1 carries only the in-plane turbulent kinetic energy, which the planar PIV
measures directly, whereas M4 relies on the full three-dimensional TKE
$k=\tfrac12\langle u'^2+v'^2+w'^2\rangle$; with $\langle w'^2\rangle$ unmeasured, M4's energy is
supervised against the in-plane surrogate $\tfrac12\langle u'^2+v'^2\rangle$---a mismatch we accept in
order to train the model. An in-plane closure is thus the natural match for an in-plane
measurement; the point is that the solver-agnostic trainer runs on experimental data ---needing
only a patched mean-velocity sample rather than a resolved DNS.

\begin{table}
  \centering
  \caption{Mean relative-$L^2$ error vs.\ DNS (averaged over the six shapes) for M1--M4 and SST
  $k$--$\omega$, in-sample (top) and leave-one-shape-out (bottom). \textbf{Bold} marks the best
  learned closure in each row; a dash marks a quantity a closure does not model. Drag is in
  table~\ref{tab:drag}.}
  \label{tab:results}
  \small
  \begin{tabular}{lccccc}
    \toprule
    quantity & M1 (local) & M2 (alg.\ $\lmix$) & M3 (learned $\lmix$) & M4 (force) & SST \\
    \midrule
    \multicolumn{6}{l}{\textit{In-sample}}\\
    $U$        & \rev{0.056} & \rev{0.038} & \rev{\textbf{0.028}} & \rev{0.038} & 0.215 \\
    $V$        & \rev{0.247} & \rev{0.212} & \rev{\textbf{0.153}} & \rev{0.192} & 0.780 \\
    $p$        & \rev{0.333} & \rev{\textbf{0.184}} & \rev{0.188} & \rev{0.221} & 0.621 \\
    $k$        & \rev{0.488} & \rev{0.235} & \rev{\textbf{0.230}} & \rev{0.302} & 0.725 \\
    $\tau_{xx}$& \rev{0.442} & \rev{0.622} & \rev{\textbf{0.415}} & --- & 0.737 \\
    $\tau_{xy}$& \rev{\textbf{0.317}} & \rev{0.364} & \rev{0.339} & --- & 0.797 \\
    $\tau_{yy}$& \rev{0.486} & \rev{\textbf{0.282}} & \rev{0.306} & --- & 0.828 \\
    $\tzz$     & --- & \rev{0.223} & \rev{\textbf{0.215}} & --- & 0.627 \\
    $F_x$      & \rev{0.720} & \rev{0.609} & \rev{0.608} & \rev{\textbf{0.579}} & 0.904 \\
    $F_y$      & \rev{0.787} & \rev{0.478} & \rev{0.448} & \rev{\textbf{0.430}} & 0.926 \\
    \midrule
    \multicolumn{6}{l}{\textit{Leave-one-shape-out}}\\
    $U$        & \rev{0.072} & \rev{0.074} & \rev{0.079} & \rev{\textbf{0.051}} & 0.215 \\
    $V$        & \rev{0.313} & \rev{0.358} & \rev{0.320} & \rev{\textbf{0.239}} & 0.780 \\
    $p$        & \rev{0.366} & \rev{0.265} & \rev{\textbf{0.257}} & \rev{0.304} & 0.621 \\
    $k$        & \rev{0.611} & \rev{0.305} & \rev{\textbf{0.295}} & \rev{0.446} & 0.725 \\
    $\tau_{xx}$& \rev{0.669} & \rev{0.694} & \rev{\textbf{0.516}} & --- & 0.737 \\
    $\tau_{xy}$& \rev{0.598} & \rev{\textbf{0.481}} & \rev{0.496} & --- & 0.797 \\
    $\tau_{yy}$& \rev{0.570} & \rev{\textbf{0.364}} & \rev{0.406} & --- & 0.828 \\
    $\tzz$     & --- & \rev{0.430} & \rev{\textbf{0.378}} & --- & 0.627 \\
    $F_x$      & \rev{0.797} & \rev{0.784} & \rev{0.692} & \rev{\textbf{0.653}} & 0.904 \\
    $F_y$      & \rev{0.828} & \rev{0.601} & \rev{0.657} & \rev{\textbf{0.543}} & 0.926 \\
    \bottomrule
  \end{tabular}
\end{table}

\begin{table}
  \centering
  \caption{Drag coefficient $C_d$ per shape for DNS, M1--M4 and SST
  $k$--$\omega$, in-sample (top) and leave-one-shape-out (bottom). \textbf{Bold} marks the closure
  closest to DNS in each row; the last row of each block is the mean relative drag error
  $\langle|\Delta C_d|/C_d^{\mathrm{DNS}}\rangle$.}
  \label{tab:drag}
  \begin{tabular}{lcccccc}
    \toprule
    shape & DNS & M1 (local) & M2 (alg.\ $\lmix$) & M3 (learned $\lmix$) & M4 (force) & SST \\
    \midrule
    \multicolumn{7}{l}{\textit{In-sample}}\\
    circle              & 1.21 & \rev{\textbf{1.22}} & \rev{1.06} & \rev{1.02} & \rev{1.40} & \rev{1.42} \\
    square              & 2.15 & \rev{1.50} & \rev{2.00} & \rev{1.82} & \rev{\textbf{2.10}} & \rev{2.48} \\
    ellipse             & 0.61 & \rev{0.86} & \rev{\textbf{0.60}} & \rev{0.57} & \rev{0.74} & \rev{0.86} \\
    triangle (long)     & 1.07 & \rev{1.20} & \rev{1.01} & \rev{\textbf{1.06}} & \rev{1.22} & \rev{1.06} \\
    triangle (equi.)    & 1.39 & \rev{1.50} & \rev{1.33} & \rev{\textbf{1.38}} & \rev{1.58} & \rev{1.58} \\
    diamond             & 2.60 & \rev{2.39} & \rev{2.37} & \rev{2.41} & \rev{2.77} & \rev{\textbf{2.72}} \\
    mean rel.\ err.     & ---  & \rev{0.167} & \rev{\textbf{0.064}} & \rev{0.076} & \rev{0.124} & \rev{0.155} \\
    \midrule
    \multicolumn{7}{l}{\textit{Leave-one-shape-out}}\\
    circle              & 1.21 & \rev{\textbf{1.23}} & \rev{0.91} & \rev{0.97} & \rev{1.19} & \rev{1.42} \\
    square              & 2.15 & \rev{1.80} & \rev{1.79} & \rev{\textbf{1.88}} & \rev{1.75} & \rev{2.48} \\
    ellipse             & 0.61 & \rev{0.76} & \rev{0.62} & \rev{\textbf{0.60}} & \rev{0.59} & \rev{0.86} \\
    triangle (long)     & 1.07 & \rev{1.29} & \rev{0.98} & \rev{0.90} & \rev{1.05} & \rev{\textbf{1.06}} \\
    triangle (equi.)    & 1.39 & \rev{1.55} & \rev{1.27} & \rev{\textbf{1.38}} & \rev{1.50} & \rev{1.58} \\
    diamond             & 2.60 & \rev{2.43} & \rev{2.17} & \rev{2.10} & \rev{2.82} & \rev{\textbf{2.72}} \\
    mean rel.\ err.     & ---  & \rev{0.137} & \rev{0.130} & \rev{0.117} & \rev{\textbf{0.070}} & \rev{0.155} \\
    \bottomrule
  \end{tabular}
\end{table}

\subsection{In-sample accuracy}\label{sec:results-insample}

\begin{figure}
  \centering
  \includegraphics[width=\textwidth]{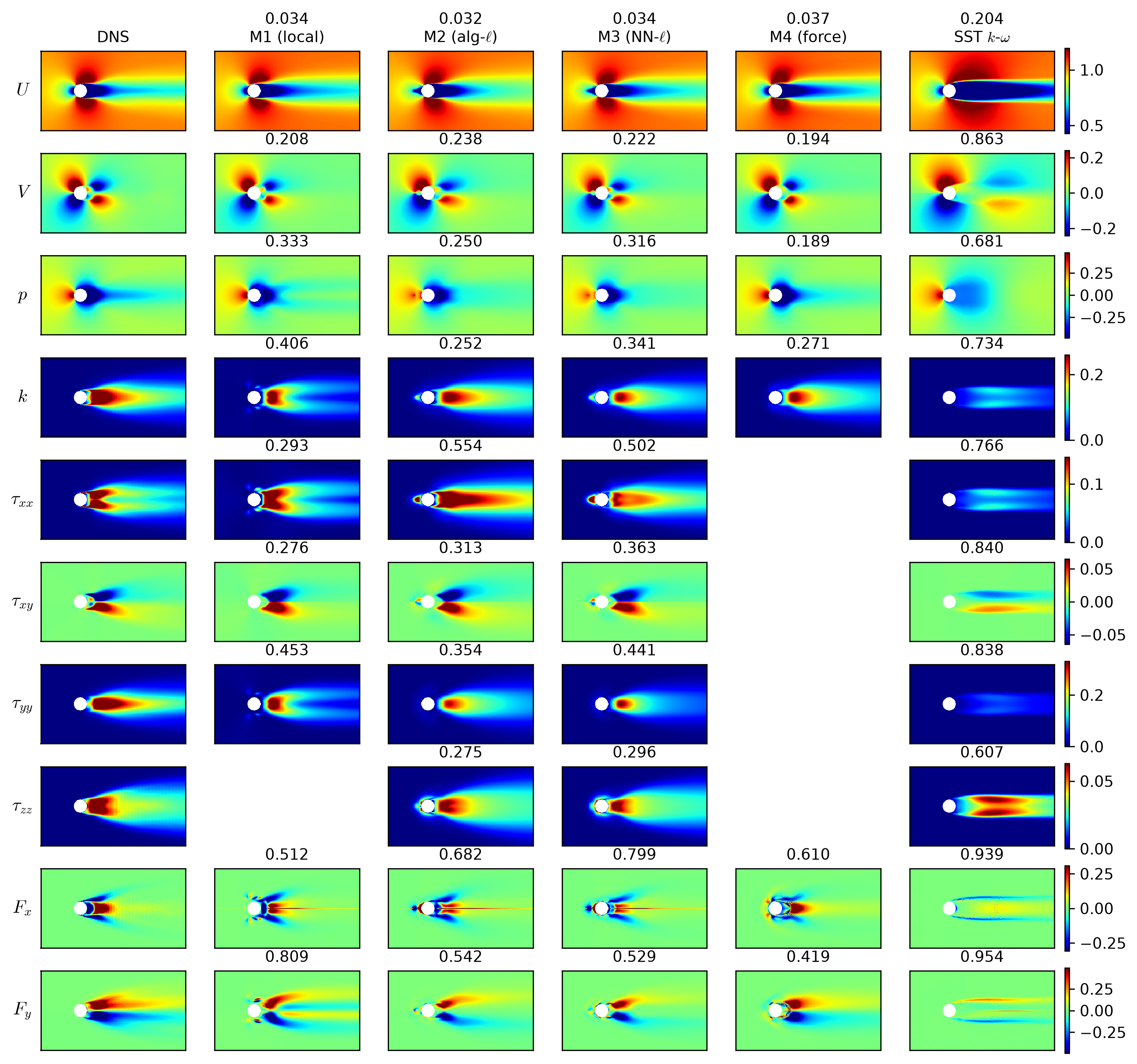}
  \caption{In-sample six-way field comparison for the circular cylinder. Columns (left to
  right): the DNS reference, then the five deployed models M1 (local), M2 (algebraic $\lmix$), M3
  (learned $\lmix$), M4 (force), and SST $k$--$\omega$. Rows (top to bottom): the mean
  streamwise and cross-stream velocity $U,V$, the pressure $p$, the turbulent kinetic energy $k$, the
  in-plane Reynolds stresses $\tau_{xx},\tau_{xy},\tau_{yy}$, the out-of-plane normal stress $\tzz$,
  and the two Reynolds-force components $F_x,F_y$.
  The number above each panel is the field's
  relative-$L^2$ error against DNS. Blank panels are quantities a model does not represent: M1 omits $\tzz$ and the force model
  M4 omits the four stresses.}
  \label{fig:insample}
\end{figure}

\begin{figure}
  \centering
  \includegraphics[width=\textwidth]{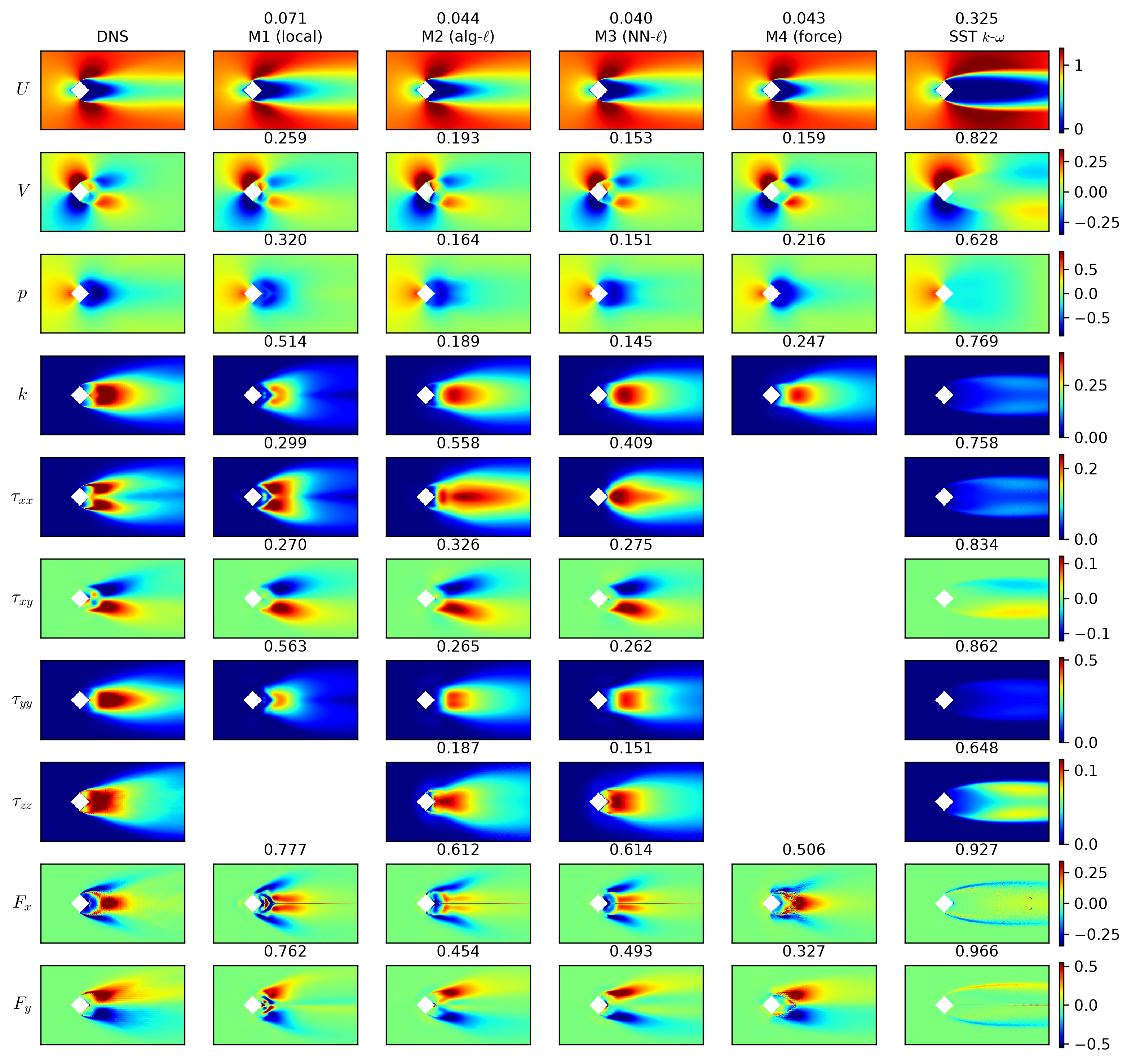}
  \caption{In-sample six-way field comparison for the diamond, a sharp-cornered bluff body whose wake
  differs markedly from the circular cylinder. Layout exactly as in figure~\ref{fig:insample}: columns are
  the DNS reference and the deployed models M1 (local), M2 (algebraic $\lmix$), M3 (learned
  $\lmix$), M4 (force), SST $k$--$\omega$; rows are the fields $U,V,p,k,\tau_{xx},\tau_{xy},\tau_{yy},
  \tzz,F_x,F_y$; and the number above each panel is that field's relative-$L^2$ error against DNS.}
  \label{fig:app-is-sq45}
\end{figure}

\begin{figure}
  \centering
  \includegraphics[width=\textwidth]{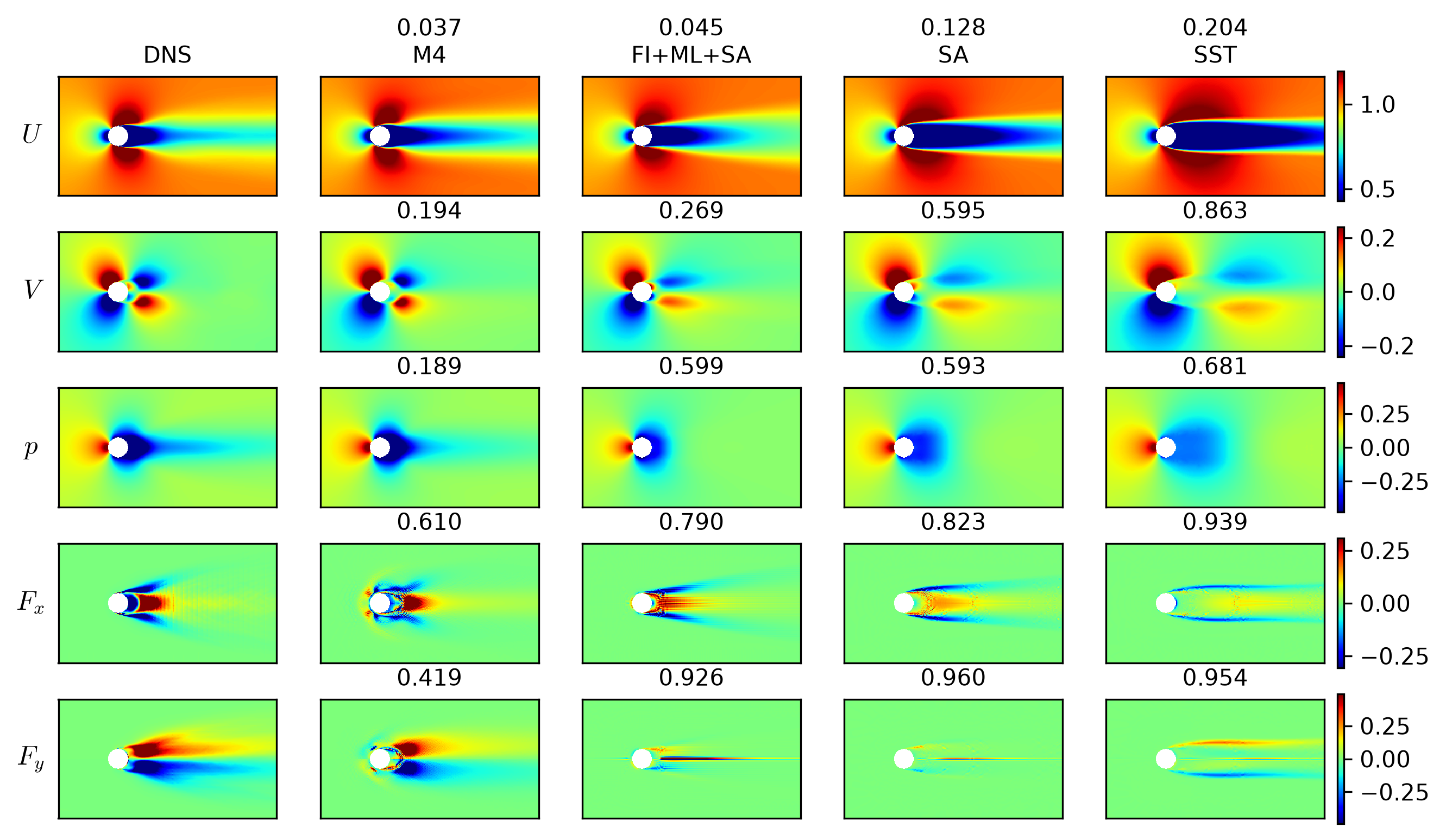}
  \caption{\rev{Comparison against the state-of-the-art data-driven closures: force model M4, in-sample on the
  circular cylinder.} Columns (left to right): the DNS reference; M4; a
  field-inversion-and-machine-learning closure that learns a corrective multiplier $\beta$ on the
  Spalart--Allmaras production term (FI+ML+SA) \citep{zhang2026}; the bare Spalart--Allmaras model (SA); and SST
  $k$--$\omega$. Rows (top to bottom): the mean velocity $U,V$, the pressure $p$, and the
  Reynolds-force components $F_x,F_y$. The number above each panel is that
  field's relative-$L^2$ error against DNS.}
  \label{fig:dafoam}
\end{figure}

The in-sample block of table~\ref{tab:results} reports the errors averaged over the six shapes; the
six-way field comparison for the cylinder is figure~\ref{fig:insample} and for the diamond
figure~\ref{fig:app-is-sq45}, with the remaining four geometries in appendix~\ref{app:fields} (the
square, figure~\ref{fig:app-is-sq0}; the ellipse, figure~\ref{fig:app-is-ell}; the long triangle,
figure~\ref{fig:app-is-trilong}; and the equilateral triangle, figure~\ref{fig:app-is-triequi}). \rev{All four learned closures yield lower shape-averaged relative $L^2$ errors
than the SST $k$--$\omega$ baseline for every field they predict. The SST errors in the two mean-velocity
components are approximately three to eight times those of the learned closures.} Among the stress closures the learned
$k$-free length scale (M3) is the most accurate overall: relative to the local
model it reduces the pressure error by \rev{$1.8\times$ ($0.188$ vs.\ $0.333$)}, velocity $V$ by
\rev{$1.6\times$}, and the turbulent kinetic energy by \rev{$2.1\times$}, and it predicts the
out-of-plane stress $\tzz$ (\rev{$0.215$}), which the two-dimensional local model cannot represent; only
\rev{the pressure and the shear and cross-stream normal stresses ($p$, $\tau_{xy}$, $\tau_{yy}$) are captured marginally better
by M2, M1 and M2, respectively} (table~\ref{tab:results}). This local deficit is structural: because M1
maps the local velocity gradients pointwise, it cannot carry the upstream history that
organizes the wake, so it reconstructs the wake incorrectly---its $k$ and stress fields show a
qualitatively different spatial pattern from DNS in figure~\ref{fig:insample}---a limitation only the
non-local transported-$k$ closures (M2--M4) remove. The
comparison between M2 and M3 isolates the length scale: the algebraic-$\lmix$ model M2 loses the
in-plane normal stress to over-isotropy (\rev{$\mathcal E(\tau_{xx})=0.622$}), which the learned $k$-free length scale
restores (\rev{$0.415$}). The forcing-vector model M4 is competitive on the mean field while modelling only
the two-component force rather than the full tensor: its mean velocity (\rev{$\mathcal E(U)=0.038$}, \rev{$\mathcal E(V)=0.192$}) trails
only M3 and \rev{matches or beats} the other two stress closures, and it predicts the Reynolds force
$\mathbf F=-\bnabla\!\cdot\tij$ itself most accurately of all (\rev{$\mathcal E(F_x)=0.579$}, \rev{$\mathcal E(F_y)=0.430$}, where SST
reaches only $0.904$ and $0.926$). The diamond is the decisive case for the force model: an unconstrained force closure locks onto a
\rev{numerically attracting but inaccurate fixed point (relative $L^2$ error in
$U$ of approximately $0.69$)}, but the \rev{regularized projection} \eqref{eq:perp} restores
\rev{an error of $0.043$}---the deployed (projected) M4 shown for the diamond in figure~\ref{fig:app-is-sq45}.
\rev{Here, attraction refers to the numerical iteration, not physical stability:
a closure-coupled fixed point need not reproduce DNS. Neural closures can introduce
multiple solutions \citep{zhang2026tangent}, while ill-conditioning can amplify
closure errors even without solution multiplicity \citep{wu2019}.} \rev{For
the drag coefficient (table~\ref{tab:drag}), the mean relative errors
of M2 and M3 are $6.4\%$ and $7.6\%$, that of M4 is $12.4\%$, and those of the local model M1 and
the SST $k$--$\omega$ baseline are comparable, at $16.7\%$ and $15.5\%$.}

It is instructive to place these closures against the established data-driven paradigm rather than
against eddy-viscosity baselines alone. Figure~\ref{fig:dafoam} compares the in-sample force model M4
on the cylinder with a field-inversion-and-machine-learning (FIML) closure of the solver-in-the-loop
type discussed in \S\ref{sec:intro}: a neural network predicts a corrective multiplier $\beta$ on the
production term of the Spalart--Allmaras (SA) model \citep{spalart1992}, with the FIML case set up as
in our earlier work \citep{zhang2026} and trained against DNS; the bare SA model is included
as the un-augmented baseline. The FIML correction
appreciably improves SA on the velocity and force fields (mean velocity $U$ from $0.128$ to $0.045$,
$V$ from $0.595$ to $0.269$; $F_x$ from $0.823$ to $0.790$), confirming that data-driven augmentation
helps. 
The solver-agnostic force model M4 performs comparably on the mean velocity fields (\rev{$\mathcal E(U)=0.037$} versus $0.045$, \rev{$\mathcal E(V)=0.194$} versus $0.269$) but is significantly more accurate on the pressure (\rev{$0.189$} versus $0.599$) and the Reynolds force components (\rev{$\mathcal E(F_x)=0.610$}, \rev{$\mathcal E(F_y)=0.419$} versus $0.790$ and $0.926$).
The gap is widest on the pressure and
the force, the quantities that enter the momentum balance directly: by learning the force as a free
field rather than a single scalar correction tethered to the SA production structure, M4 captures the
near-wake forcing that a production-term multiplier cannot.

\subsection{Generalization: leave-one-shape-out}\label{sec:results-loso}

\begin{figure}
  \centering
  \includegraphics[width=\textwidth]{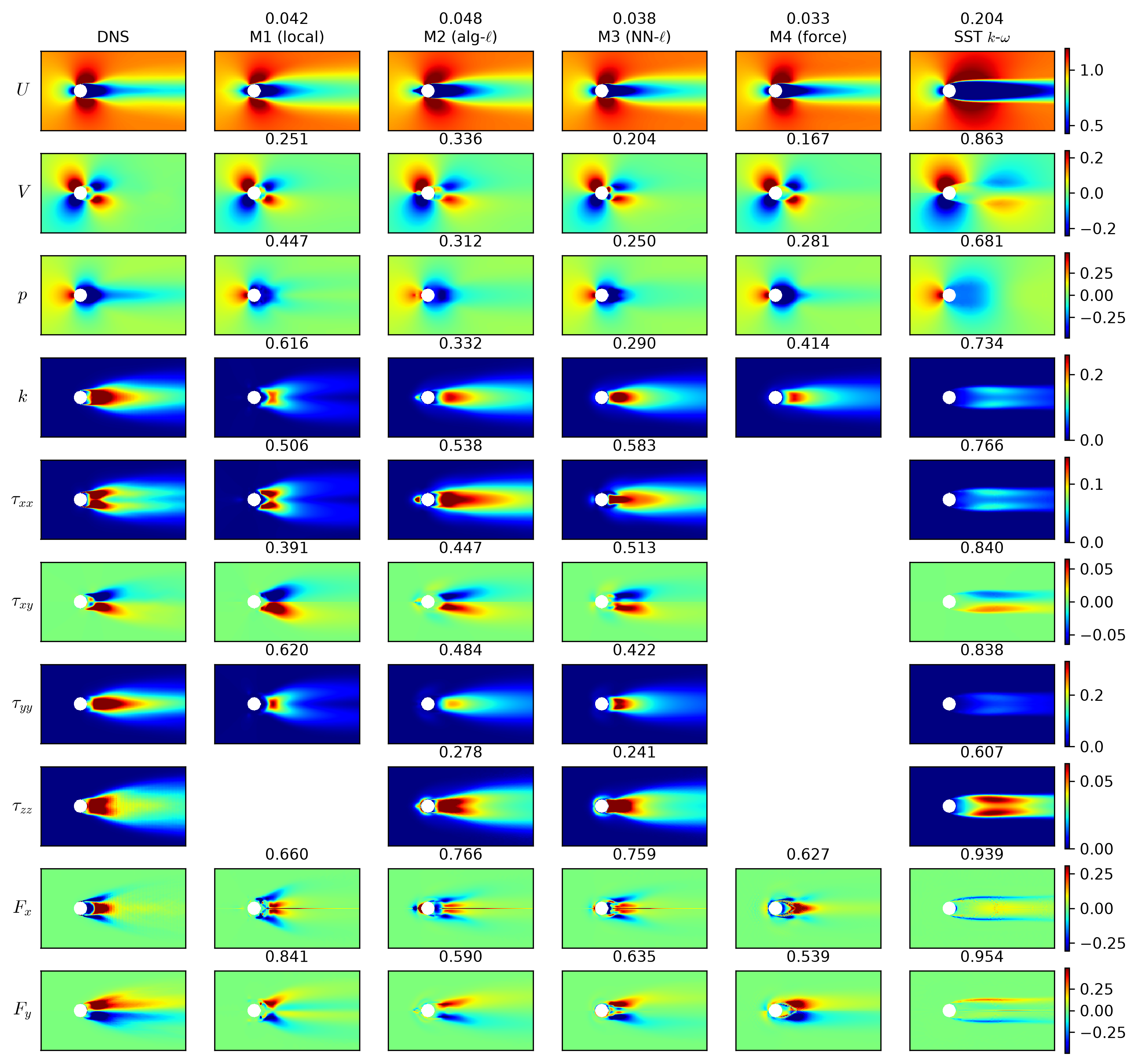}
  \caption{Leave-one-shape-out six-way field comparison for the held-out circular cylinder (each
  model trained on the other five shapes and deployed on this unseen one). Layout exactly as in
  figure~\ref{fig:insample}: columns are the DNS reference and the deployed models M1 (local), M2
  (algebraic $\lmix$), M3 (learned $\lmix$), M4 (force), SST $k$--$\omega$; rows are the fields
  $U,V,p,k,\tau_{xx},\tau_{xy},\tau_{yy},\tzz,F_x,F_y$; and the number above each panel is that
  field's relative-$L^2$ error against DNS.}
  \label{fig:loso}
\end{figure}

\begin{figure}
  \centering
  \includegraphics[width=\textwidth]{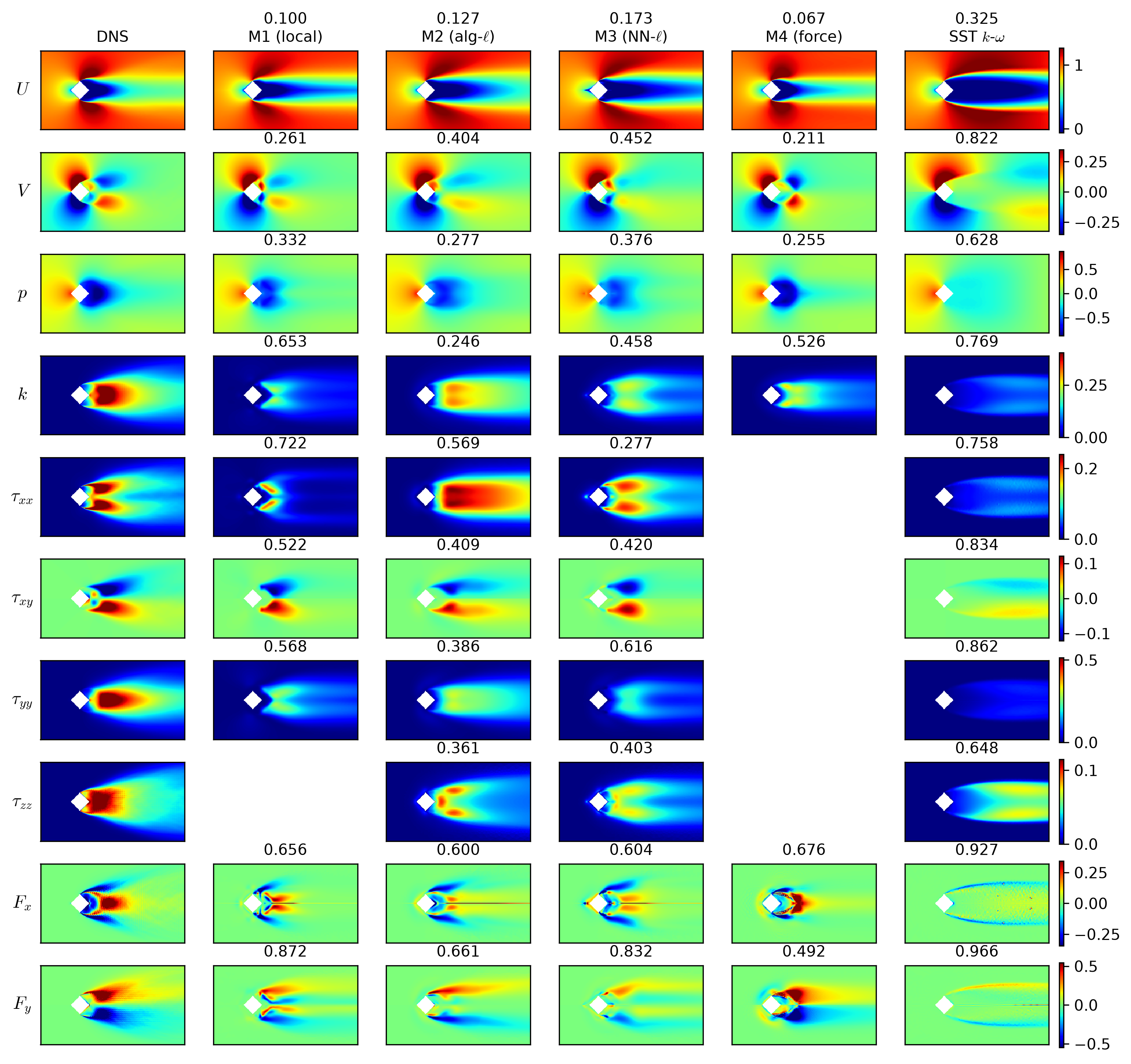}
  \caption{Leave-one-shape-out six-way field comparison for the held-out diamond (trained on the
  other five shapes). Layout as in figure~\ref{fig:insample}: columns are the DNS reference and the
  models M1--M4 and SST $k$--$\omega$; rows are $U,V,p,k,\tau_{xx},\tau_{xy},\tau_{yy},\tzz,F_x,F_y$;
  the number above each panel is that field's relative-$L^2$ error against DNS.}
  \label{fig:app-lo-sq45}
\end{figure}
Generalization is the decisive test, assessed from the LOSO block of table~\ref{tab:results}, the
drag table~\ref{tab:drag}, the held-out circle of figure~\ref{fig:loso}, the held-out diamond of
figure~\ref{fig:app-lo-sq45}, and the remaining \rev{four}
held-out shapes in appendix~\ref{app:fields} (the square, figure~\ref{fig:app-lo-sq0}; \rev{the ellipse, figure~\ref{fig:app-lo-ell};} the long
triangle, figure~\ref{fig:app-lo-trilong}; and the equilateral triangle,
figure~\ref{fig:app-lo-triequi}).
Figure~\ref{fig:loso} shows the held-out circle: every learned closure stays finite, but the
non-local models track DNS markedly better than the local one on the wake-defining fields---M1 is the
least accurate there on the pressure (\rev{$0.447$}), the energy (\rev{$0.616$}), and the cross-stream normal
stress (\rev{$\mathcal E(\tau_{yy})=0.620$}), while M4 recovers the mean velocity best (\rev{$\mathcal E(U)=0.033$}, \rev{$\mathcal E(V)=0.167$}).

Averaged over all six held-out shapes (table~\ref{tab:results}) this hardens into a clear ordering.
The local closure M1 generalizes worst: \rev{its energy, shear-stress and cross-stream normal-stress errors
are the largest of the four (mean $\mathcal E(k)=0.611$, $\mathcal E(\tau_{xy})=0.598$, $\mathcal E(\tau_{yy})=0.570$)}, the price of a purely local map
extrapolated off the training manifold. Among the stress closures the learned length scale M3 is the
most accurate on \rev{the streamwise normal stress, the energy and the pressure (lowest $\mathcal E(\tau_{xx})=0.516$, $\mathcal E(k)=0.295$, $\mathcal E(p)=0.257$, and out-of-plane
$\mathcal E(\tzz)=0.378$), the algebraic-$\lmix$ model M2 edging it on the shear stress ($0.481$ vs.\ $0.496$)
and the cross-stream normal stress ($0.364$ vs.\ $0.406$)}. The forcing-vector model M4 is the
strongest generalizer on the mean flow: it is the best of the four on the mean velocity (\rev{$\mathcal E(U)=0.051$}
vs.\ M3's \rev{$0.079$}, \rev{$\mathcal E(V)=0.239$} vs.\ \rev{$0.320$}), predicts the held-out Reynolds force best (\rev{$\mathcal E(F_x)=0.653$},
\rev{$\mathcal E(F_y)=0.543$} vs.\ SST's $0.904$, $0.926$), and recovers the drag most accurately, a leave-one-shape-out mean error of \rev{$7.0\%$} against its
own in-sample \rev{$12\%$}, the stress closures' \rev{$12$--$14\%$, and SST's $15.5\%$ (within $\sim$2\% on the cylinder and the long triangle},
table~\ref{tab:drag}). \rev{Drag is an integrated surface quantity, dominated
here by pressure: $C_d\simeq-2\int_{\partial B}P n_x\,\mathrm ds$ in the
present nondimensionalization, with $\mathbf n$ directed from the body into the
fluid. Surface integration permits cancellation of pressure errors. Thus the
$7.0\%$ drag error does not imply comparable local wake accuracy, as the larger
force and stress errors demonstrate.} The hardest geometry is
the ellipse (figure~\ref{fig:app-lo-ell})---the streamlined, smooth-body separation least like the
bluffer training shapes---on
which the stress closures extrapolate worst (M2 reaches \rev{$\mathcal E(\tau_{xx})=0.969$} there), while the
sharp-cornered square and diamond are the hardest on the mean velocity. In short, M4 generalizes
best on the mean flow, force, and drag, and M3 best on the stress field, both ahead of the local
closure\rev{, and all four ahead of the SST baseline on every field quantity}.

Although the present study fixes $\Rey=10^4$, the generalization is expected to extend in Reynolds
number as well. Flows past a circular cylinder settle into a common, weakly Reynolds-dependent mean-flow regime
above $\Rey\approx5000$ until the emergence of the drag crisis after $\Rey\approx200,000$---a pattern our prior cylinder study documents across Reynolds number up to $\Rey=140,000$
\citep{zhang2026}. Therefore, we expect that a closure that transfers across shapes at a single Reynolds number will also
transfer across Reynolds number in this regime. The framework therefore targets generalization in
both geometry and Reynolds number, with the present six-shape study isolating the harder geometric axis.

\section{Conclusion}\label{sec:conclusion}

We have trained turbulence closures entirely inside a physics-informed neural network---imposing the
RANS residual by automatic differentiation, with no external solver, mesh, or adjoint in the inverse
loop---and used the same solver-agnostic framework to pursue two modelling targets on equal footing:
the Reynolds stress, in a tensor basis (three closures of increasing structure---a local map,
a non-local transported-$k$ model that also recovers the out-of-plane normal stress, and the same
with a learned length scale), and the Reynolds force it exerts on the mean flow (a
structure-preserving forcing-vector model, M4). Across six distinct bluff-body wakes, deployed frozen
in a standard finite-element solver, all four closures beat the SST $k$--$\omega$ baseline \rev{on every field quantity} in both the
in-sample and generalization settings\rev{, by a wide margin on the mean velocity}. \rev{The advantage in drag is
smaller: the mean drag error of SST ($15.5\%$) is comparable to that of the local closure in-sample and about
twice that of the force closure on held-out shapes.} The learned-length-scale stress closure
(M3) is the most accurate on the stress fields; the force closure (M4) \rev{beats it on the
mean velocity and the drag on held-out shapes}, uniquely predicts the momentum force, and generalizes best of the four on
the mean flow and drag.
Because no forward solve runs in the training loop, each closure
trains in minutes on a single GPU---orders of magnitude cheaper than a comparable adjoint
optimization. Stable a-posteriori deployment---the bar most data-driven closures fail to clear---is
secured \rev{primarily by input-gradient smoothing}.
Because the trainer needs no resolved DNS---only enough data to anchor the mean field---it also learns directly from patched, in-plane
experimental PIV measurements.

Several limitations bound the present claims and frame future work.
The study is at a single Reynolds number. We expect the closure to
transfer across Reynolds number as well, but confirming this is the natural next test. 
The Reynolds-stress error, while well below SST, remains modest in absolute terms,
and the ellipse---the smooth-body separation least like the training set, where the closures
extrapolate worst---is the present weak point. Finally, the half-domain symmetric formulation
restricts the method to flows with a symmetric mean; relaxing it would broaden applicability to
time-dependent vortex shedding or lifting configurations. Addressing these would strengthen the case that PINN-trained,
solver-agnostic stress and force closures are a practical route to geometry-generalizable RANS modelling.

\backsection[Declaration of Interests]{The authors report no conflict of interest.}

\backsection[Data availability statement]{\rev{The code for PINN training, finite-element
deployment, and postprocessing is openly available at
\url{https://github.com/zhangzhen117/Bluff_body_closure_PINN}. The DNS and PIV data are
available from the corresponding author on reasonable request.}}

\backsection[Acknowledgements]{
This research was supported by the Defense Advanced Research Projects Agency (DARPA) under the Automated Prediction Aided by Quantized Simulators (APAQuS) program, Grant No.~HR00112490526.
\emph{Use of AI tools.} \rev{The authors used Claude Opus~4.8 (Anthropic) during
June--July 2026 to write and debug the training and finite-element deployment code, and to assist with language polishing of the manuscript text.
All AI-assisted code and text were reviewed and validated by the authors, who take full responsibility.}}

\bibliographystyle{jfm}
\bibliography{refs}

\appendix

\section{PIV--DNS comparisons for the diamond and long triangle}\label{app:piv}

\begin{figure}
  \centering
  \includegraphics[width=0.8\textwidth]{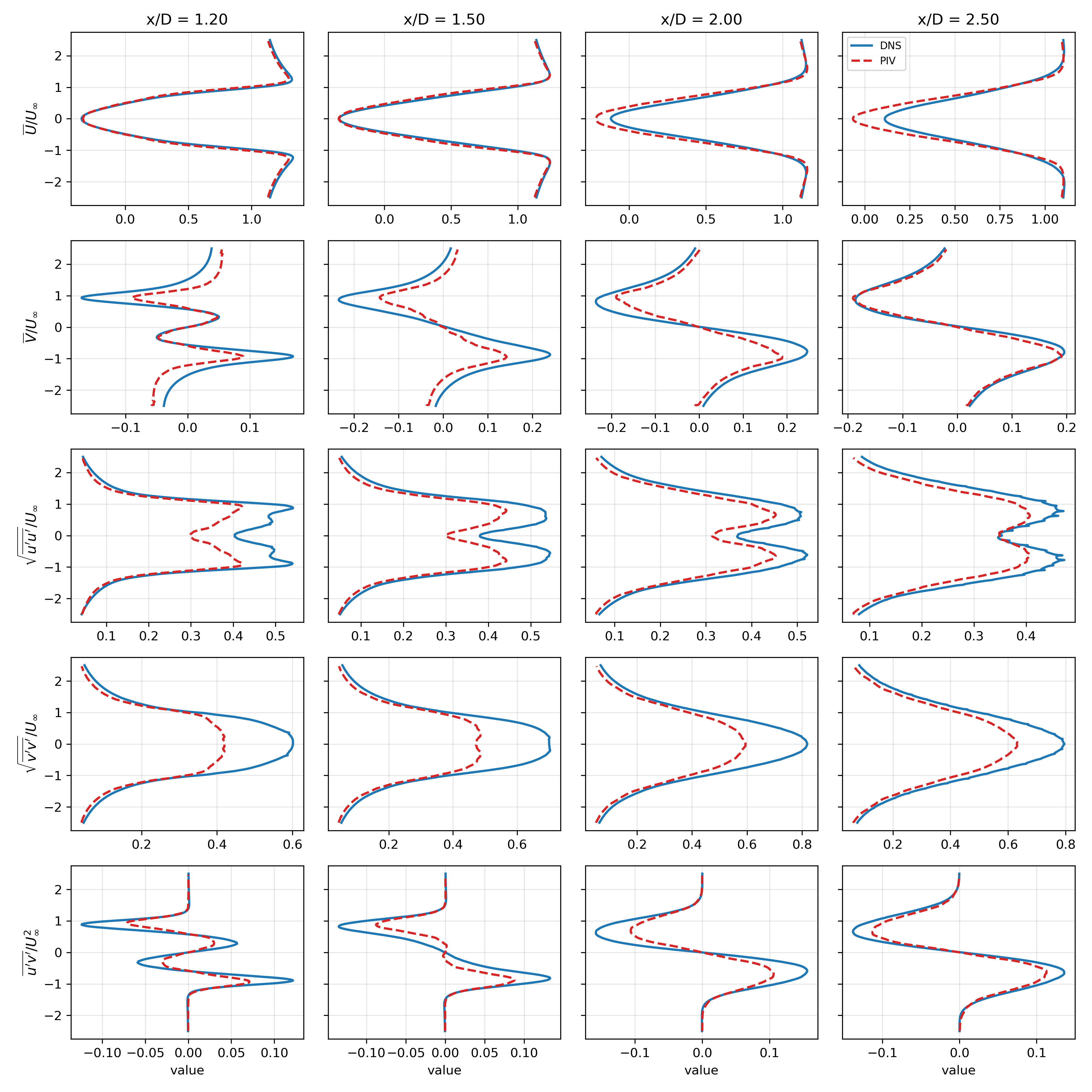}
  \caption{PIV versus DNS for the diamond at $\Rey=10^4$: wall-normal profiles (profiles as in
  figure~\ref{fig:piv-circ}).}
  \label{fig:app-piv-diam-prof}
\end{figure}

\begin{figure}
  \centering
  \includegraphics[width=0.85\textwidth]{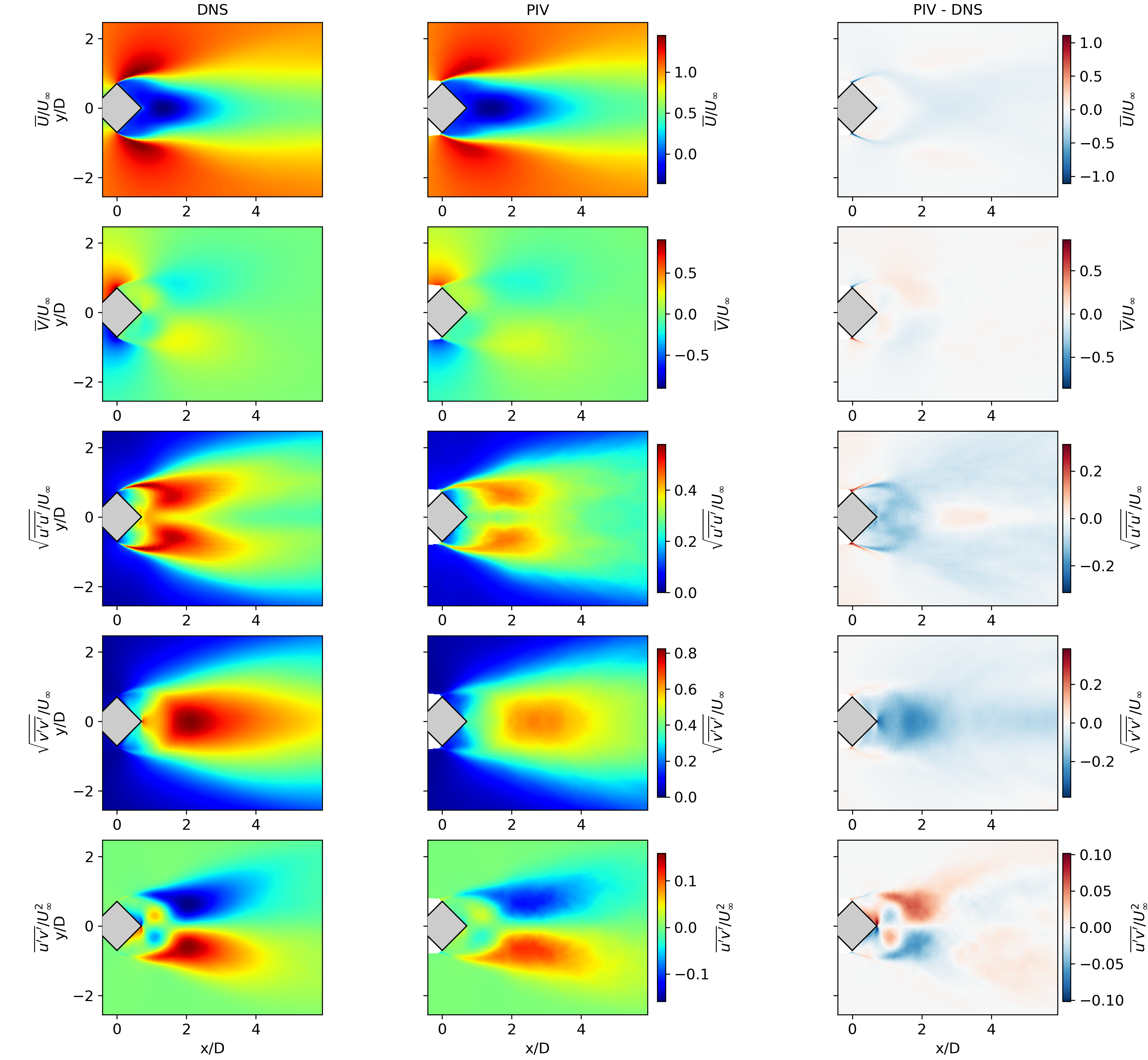}
  \caption{PIV versus DNS for the diamond at $\Rey=10^4$: full-field comparison (columns DNS, PIV,
  PIV$-$DNS).}
  \label{fig:app-piv-diam}
\end{figure}

\begin{figure}
  \centering
  \includegraphics[width=0.8\textwidth]{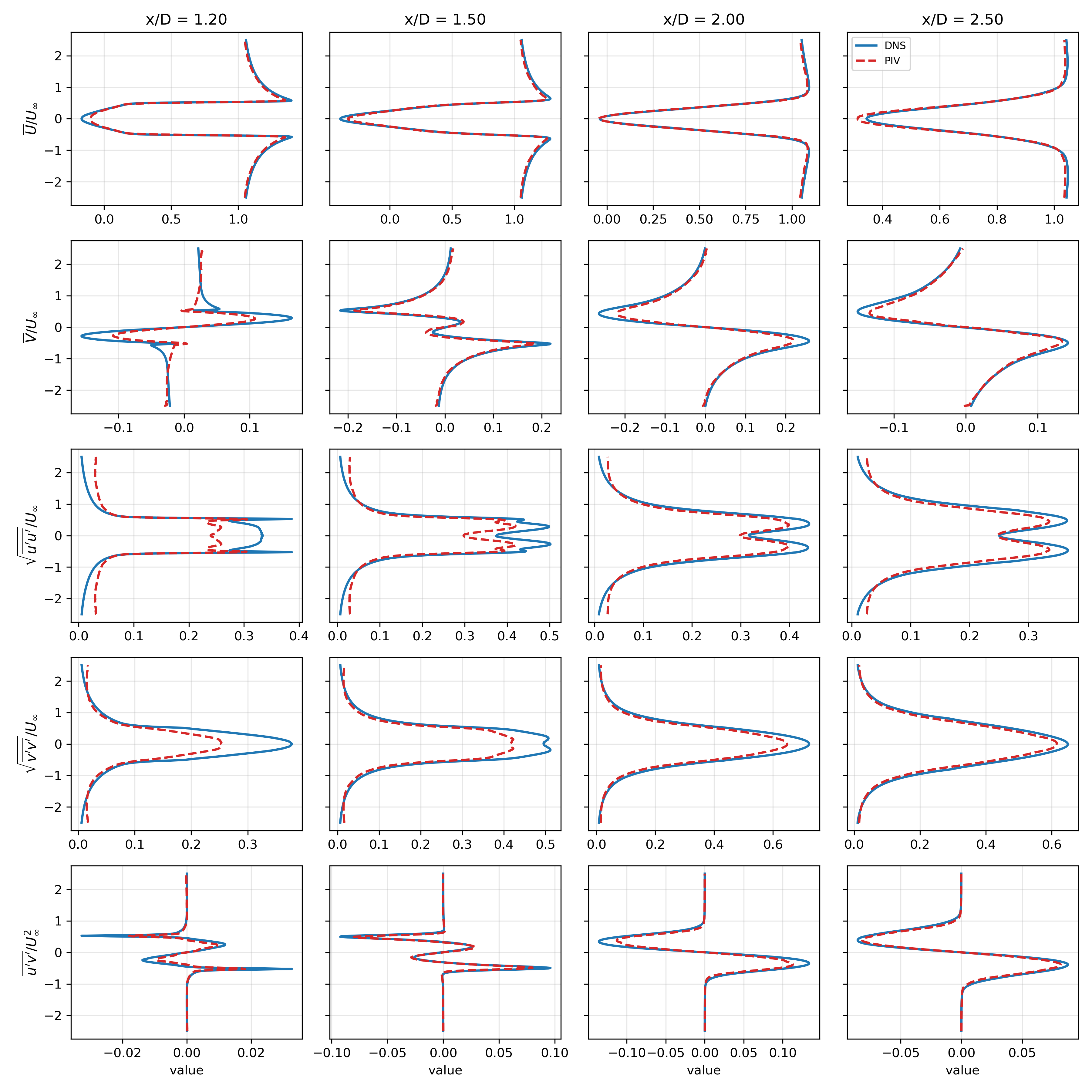}
  \caption{PIV versus DNS for the long triangle at $\Rey=10^4$: wall-normal profiles (profiles as in
  figure~\ref{fig:piv-circ}).}
  \label{fig:app-piv-tri-prof}
\end{figure}

\begin{figure}
  \centering
  \includegraphics[width=0.85\textwidth]{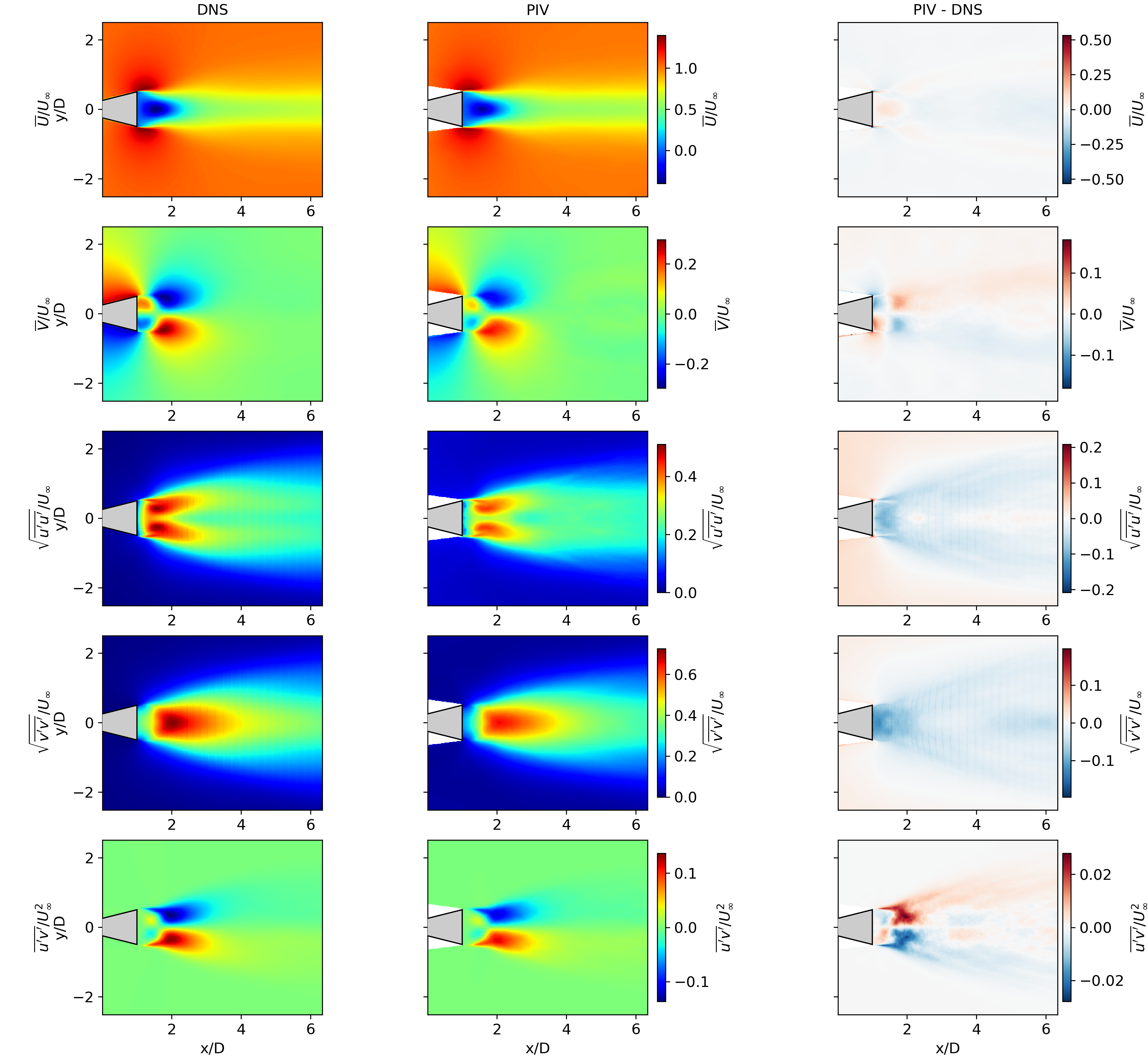}
  \caption{PIV versus DNS for the long triangle at $\Rey=10^4$: full-field comparison (columns DNS, PIV, PIV$-$DNS).}
  \label{fig:app-piv-tri}
\end{figure}

The PIV--DNS comparison for the circular cylinder is in the main text
(figures~\ref{fig:piv-circ},~\ref{fig:piv-circ-field}). The corresponding comparisons for the other
two measured shapes---the diamond (figures~\ref{fig:app-piv-diam-prof},~\ref{fig:app-piv-diam}) and
the long triangle (figures~\ref{fig:app-piv-tri-prof},~\ref{fig:app-piv-tri})---are collected here,
each shown as wake profiles and then as a full-field comparison, exactly as in
figures~\ref{fig:piv-circ} and~\ref{fig:piv-circ-field}: the profiles are taken at
$x/D=1.2,\,1.5,\,2.0,\,2.5$ with DNS solid and
PIV dashed, and the field columns are DNS, PIV, and the PIV$-$DNS difference, over the rows
$U/U_\infty$, $V/U_\infty$, $\sqrt{\langle u'^2\rangle}/U_\infty$, $\sqrt{\langle v'^2\rangle}/U_\infty$,
$\langle u'v'\rangle/U_\infty^2$. The agreement shows further evidence of the accuracy of both DNS and PIV on different shapes.

\clearpage

\clearpage

\section{\rev{Closure reconstruction and parameterization}}\label{app:closure-training}\label{app:closure-constraints}

\rev{For the three-dimensional stress state of M2/M3, the component reconstruction is as follows.}
Writing the deviatoric in-plane block
$\mathbf B=-2\nut\Smat+g_2\Tw$, with $B_{11}=-2\nut a+g_2(-2bw)$ and $B_{12}=-2\nut b+g_2(2aw)$, the
stress components are
\begin{equation}
  \tau_{xx}=\tfrac23 k+\tfrac{\Gz}{6}+B_{11},\quad
  \tau_{yy}=\tfrac23 k+\tfrac{\Gz}{6}-B_{11},\quad
  \tau_{xy}=B_{12},\quad
  \tzz=\tfrac23 k-\tfrac{\Gz}{3}.
  \label{eq:components}
\end{equation}
The stress is kept realizable: a physical stress is positive-semi-definite, and the block-diagonal $3\times3$
state requires $\tzz\ge0$ and the in-plane block PSD. With the in-plane isotropic level
$\sigma=\tfrac23 k+\tfrac{\Gz}{6}$ and the deviatoric eigenvalue magnitude
$\lambda=\sqrt{B_{11}^2+B_{12}^2}$ we cap
$\mathbf B\leftarrow\mathbf B\,\sigma\tanh(\lambda/\sigma)/\lambda$ (enforcing $\lambda\le\sigma$)
and set $\Gz=2k\tanh(\cdot)\in[-2k,2k]$; together these give $\sigma\ge k/3>0$ and $\tzz\ge0$, so the
full normal-stress state is realizable; the cap is smooth and exact in the limit. For the momentum
coupling, the isotropic and $\Tz$ parts are proportional to the identity within the in-plane block,
so their divergence is a pure gradient absorbed into a modified pressure
$P^*=P+\tfrac23 k+\tfrac{\Gz}{6}$; the mean velocity responds only to the anisotropic in-plane
stress, and $\tzz$ does not enter the two-dimensional balance but is a supervised, diagnostic output.

\rev{For M1, $k=\tfrac12\langle u'^2+v'^2\rangle$ is the in-plane kinetic energy,
$\nut=\mathrm{softplus}(\cdot)$ is a non-negative network output, and the coefficients
are capped using the in-plane isotropic level $\sigma=k$ and tapered near the wall by $\varphi(d)=\tanh(d/d_0)$. M2--M4 use the full turbulent
kinetic energy $k=\tfrac12\langle u'^2+v'^2+w'^2\rangle$ and the bounded multiplier
$m=\exp(\beta\tanh(\cdot))$, with $\beta=1.4$ (approximately $0.25\le m\le4$).
In \eqref{eq:nut}, $\varepsilon=10^{-6}$ regularizes the square root. The constants are
$\Cmu=0.09$ and $\kvk=0.41$, with $\lmix^{\max}=6.0$.}

\rev{For M3/M4, the bounded length-scale transformation $\mathcal H$ in
\eqref{eq:length-map} is}
\begin{equation}
  \lmix = \mathrm{clip}\!\Big(\kvk\,d\cdot e^{\,\beta\tanh(\mathcal{C}'(\mathbf q_7))},\,
  \lmix^{\min},\,\lmix^{\max}\Big),
  \label{eq:learnedell}
\end{equation}
\rev{with $\beta=1.4$, $\lmix^{\min}=10^{-3}$, and $\lmix^{\max}=6.0$.
The length-scale network receives $\mathbf q_7$ only; excluding $k$ preserves the
monotonic dependence of $k^{3/2}/\lmix$ on $k$ at fixed kinematic features.}

\clearpage
\section{\rev{PINN setup, training fields and ablations}}\label{app:pinn}

\begin{figure}
  \centering
  \includegraphics[width=\textwidth]{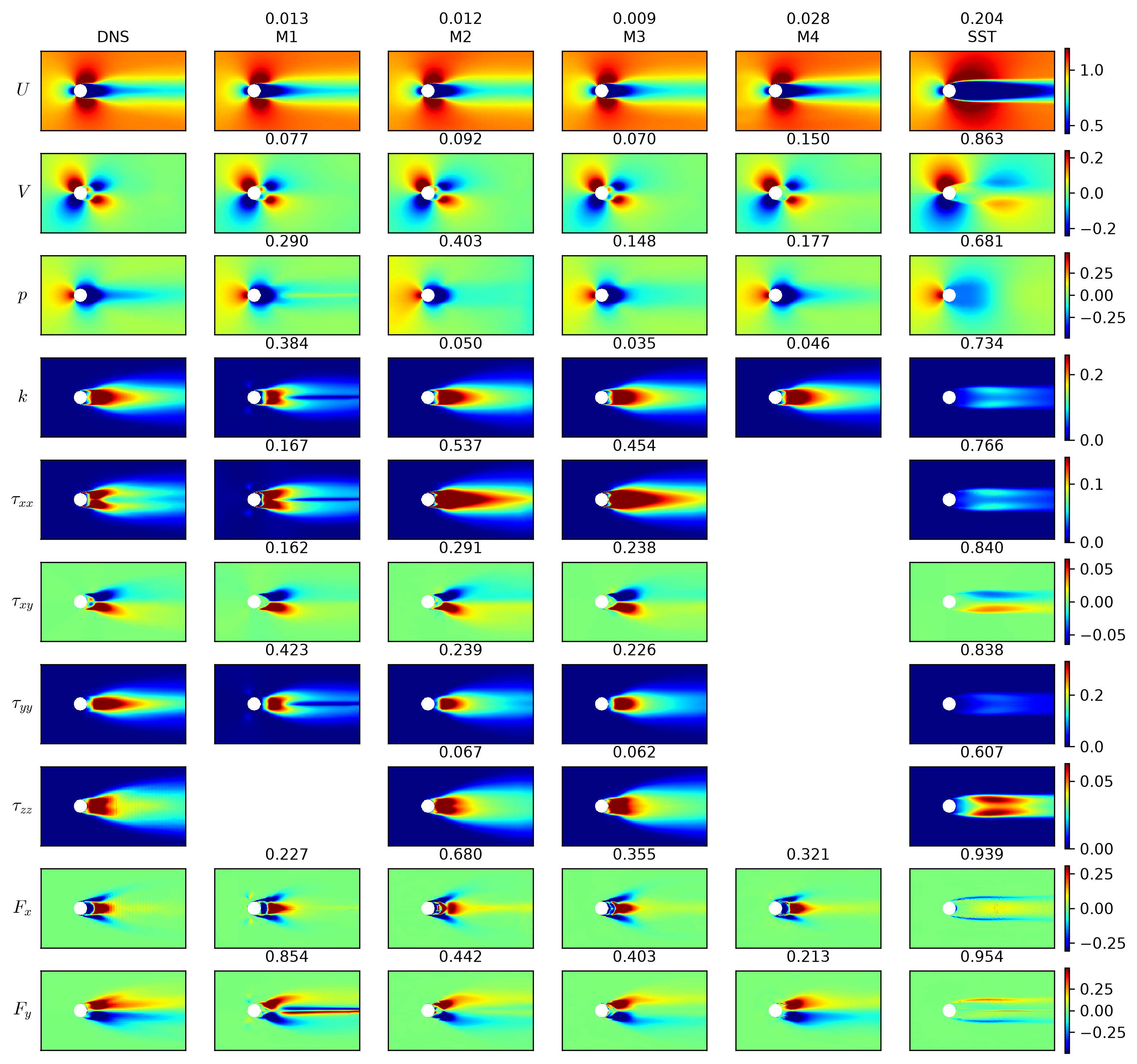}
  \caption{Supervised-PINN field comparison on the in-sample cylinder---each closure's own
  converged field, before FEM deployment. Columns: DNS, M1--M4, SST $k$--$\omega$; rows: $U,V,p,k,
  \tau_{xx},\tau_{xy},\tau_{yy},\tau_{zz},F_x,F_y$; the number above each panel is that field's
  relative-$L^2$ error against DNS. Blank panels are quantities a closure does not represent. Compare
  the FEM deployment of the same closures in figure~\ref{fig:insample}.}
  \label{fig:exp-train}
\end{figure}

This appendix collects the supervised-PINN fields---each closure read out directly from its
converged PINN, before FEM deployment---underlying the deployed results of \S\ref{sec:results}, and
the ablation of the PINN training recipe, all on the in-sample cylinder.
Figure~\ref{fig:exp-train} shows each closure's own converged field; compare the a-posteriori
(FEM) deployment in figure~\ref{fig:insample}.

\subsection{\rev{Loss weighting}}\label{app:loss-weighting}

\rev{Within each geometry, the PDE, data, and boundary losses in \eqref{eq:Lgroups}
contain adaptive weights $w_j(t)$. These are updated periodically using the gradient-norm
balancing procedure of \citet{wang2021grad}, so that the weighted terms contribute
comparable parameter-gradient magnitudes. The Lipschitz coefficient is fixed at
$\lambda_{\mathrm{lip}}=0.03$. The following expressions apply to each geometry;
the geometry index is omitted for clarity.}

The boundary conditions are penalized as
\begin{equation}
  \mathcal L_{\mathrm{bc}}
  = w_{b}\big(\mathcal L_{\mathrm{bc}}^{\bU}
  + \mathcal L_{\mathrm{bc}}^{k}\big),
  \label{eq:Lbc}
\end{equation}
imposing uniform inflow $\bU=(1,0)$, zero-pressure outflow, far-field symmetry
($V=0,\ \partial_y U=0$) and no-slip on the body ($\mathcal L_{\mathrm{bc}}^{\bU}$); and the homogeneous
$k=0$ on the body and inlet for the non-local methods ($\mathcal L_{\mathrm{bc}}^{k}$).

\rev{The shared Lipschitz regularizer is defined in \eqref{eq:Llip} and enters
\eqref{eq:loss} once, outside the sum of per-geometry objectives. Its average is taken
over feature samples from all training shapes.}

\subsection{\rev{Pseudo-time stepping}}\label{app:training-residuals}

\rev{The spatial residuals are defined in \eqref{eq:residuals}, with the
transported-$k$ equation \eqref{eq:keq} used for M2--M4.
Momentum and $k$ residuals are relaxed in pseudo-time, while continuity is unchanged,}
following the time-stepping-oriented training of \citet{cao2023tsonn} and the pseudo-time analysis
of \citet{wang2026pseudo}: at pseudo-time step $n$, each residual gains the
increment of its primary unknown $\phi_i\in\{U,V,k\}$ between the current network parameters
$\theta^{n}$ and the previous, frozen iterate $\theta^{n-1}$,
\begin{equation}
  \tilde r_i = \frac{\phi_i^{\,n}-\phi_i^{\,n-1}}{\Delta t}+r_i,
  \qquad (\phi_u,\phi_v,\phi_k)=(U,V,k),
  \quad \phi_i^{\,n}=\phi_i(\,\cdot\,;\theta^{n}),
  \label{eq:pseudo}
\end{equation}
\rev{These relaxed residuals enter the PDE loss \eqref{eq:Lpde}.
The $k$ term is present only for M2--M4. At convergence,
$\phi_i^{\,n}-\phi_i^{\,n-1}\to0$, recovering the steady spatial residuals.}

\subsection{\rev{Network and training setup}}\label{app:training-setup}

\rev{Table~\ref{tab:training-configuration} lists the network architectures, sampling,
optimization settings, training protocols, and computational cost.}

\begin{table}
\centering
\begingroup\color{black}
\caption{\rev{Network architectures and training configuration. Widths and depths refer to
hidden layers; the modified-MLP architecture follows \citet{wang2021grad}.}}
\label{tab:training-configuration}
\footnotesize
\begin{tabular}{@{}p{0.29\textwidth}p{0.65\textwidth}@{}}
\toprule
Item & Configuration\\
\midrule
Solution network $\mathcal N_s$ & Modified MLP; four layers of width 64; one network per geometry\\
Closure network $\mathcal C$ & Two layers of width 24; shared across training geometries\\
Auxiliary network $\mathcal C'$ & Two layers of width 16; shared; M3/M4 only\\
Activation & $\tanh$\\
Training domain & $[-10,15]\times[-10,10]$\\
Samples per geometry & 2048 collocation points and 2048 DNS data points, resampled each iteration\\
Optimizer and steps & \textsc{soap} \citep{vyas2024soap}; $50\,000$ steps\\
Learning-rate schedule & $3\times10^{-3}$; 2000-step warm-up; exponential decay by 0.9 every 4000 steps\\
In-sample protocol & Train and deploy on the same shape\\
Leave-one-shape-out protocol & Train on five shapes; deploy on the held-out sixth\\
Training hardware and time & One NVIDIA L40S GPU; approximately 15 min in-sample or 67 min for five training shapes\\
Deployment & Retain $\mathcal C$ and, for M3/M4, $\mathcal C'$; approximately 3 min on CPU\\
\bottomrule
\end{tabular}
\endgroup
\end{table}

\rev{At each iteration, 2048 collocation points per geometry are drawn randomly
with replacement from a fixed candidate pool. The pool combines $400\,000$ uniform
candidates in the full domain with $400\,000$ in $[-3,7]\times[-3,3]$; points
inside the body are rejected. The second region increases sampling density near
the body and in the wake.}

\subsection{\rev{Inverse-training ablation}}\label{app:ablation}

\rev{We jointly retrain the M3 solution and closure
networks on the in-sample cylinder for $50\,000$ steps, using seeds 42--44 and identical
data and sampling settings. Table~\ref{tab:inverse-ablation} removes each training
ingredient individually; vanilla training removes all four. The full recipe gives
the lowest mean PINN velocity error, $0.023$, compared with $0.099$ for vanilla training.
This experiment assesses inverse closure identification directly.}

\rev{The ingredients contribute unequally to the inverse problem. Replacing \textsc{soap} by
Adam has the largest individual effect on the PINN velocity error, which triples to $0.069$ and
varies strongly between seeds ($\pm0.031$); the second-order preconditioning is therefore the
main source of both accuracy and reproducibility. Removing gradient-norm weighting raises the
error to $0.034$, while the plain MLP ($0.028$) and the omission of pseudo-time stepping
($0.025$) have smaller effects on the trained field. Removing all four ingredients compounds
these losses: the vanilla PINN error is four times that of the full recipe.

The deployed errors show that a more accurate inverse solution yields a more accurate closure.
The full recipe gives the lowest FEM errors in $U$, $V$, $\tau_{xx}$, $\tau_{xy}$, $\tau_{zz}$
and $F_x$, and the lowest drag error ($13.6\%$, against $15$--$28\%$ for the variants). Gradient-norm
weighting matters most after deployment: without it, the errors in $U$, $p$ and $C_D$ are the
largest and vary most between seeds, consistent with unbalanced loss terms leading to an
inconsistently identified closure. Pseudo-time stepping changes the trained field little.
The vanilla closure is not uniformly worse: its errors in $p$, $k$,
$\tau_{yy}$ and $F_y$ are comparable to or below those of the full recipe, while its
streamwise normal-stress error ($0.693$) is the largest in the table and its $\tau_{zz}$ error
($0.406$) is nearly twice that of the full recipe.
A poorly converged inverse solution thus distributes the closure error unevenly across
components, and agreement in a few deployed fields does not indicate a correctly identified
closure.}

\begin{table}
  \centering
  \begingroup\color{black}
  \caption{\rev{Inverse-training ablation for M3 on the cylinder: mean $\pm$ population
  standard deviation over three seeds. The first row is the PINN velocity-vector
  relative-$L^2$ error; subsequent field rows are FEM relative-$L^2$ errors against DNS,
  followed by the relative drag error. PT denotes pseudo-time stepping; vanilla removes
  all four training ingredients.}}
  \label{tab:inverse-ablation}
  \footnotesize
  \setlength{\tabcolsep}{1.5pt}
  \begin{tabular}{@{}lcccccc@{}}
    \toprule
    Error & Full & Adam & No PT & Plain MLP & No grad-norm & Vanilla\\
    \midrule
    PINN $\bU$ & $0.023 \pm 0.002$ & $0.069 \pm 0.031$ & $0.025 \pm 0.001$ & $0.028 \pm 0.009$ & $0.034 \pm 0.005$ & $0.099 \pm 0.006$\\
    \midrule
    $U$ & $0.028 \pm 0.005$ & $0.038 \pm 0.001$ & $0.032 \pm 0.001$ & $0.034 \pm 0.001$ & $0.051 \pm 0.030$ & $0.040 \pm 0.010$\\
    $V$ & $0.153 \pm 0.040$ & $0.223 \pm 0.023$ & $0.183 \pm 0.016$ & $0.213 \pm 0.004$ & $0.274 \pm 0.127$ & $0.211 \pm 0.047$\\
    $p$ & $0.247 \pm 0.041$ & $0.261 \pm 0.018$ & $0.270 \pm 0.013$ & $0.289 \pm 0.006$ & $0.401 \pm 0.154$ & $0.239 \pm 0.051$\\
    $k$ & $0.264 \pm 0.021$ & $0.275 \pm 0.009$ & $0.299 \pm 0.015$ & $0.289 \pm 0.014$ & $0.380 \pm 0.115$ & $0.247 \pm 0.052$\\
    $\tau_{xx}$ & $0.430 \pm 0.044$ & $0.463 \pm 0.037$ & $0.487 \pm 0.042$ & $0.462 \pm 0.031$ & $0.554 \pm 0.063$ & $0.693 \pm 0.145$\\
    $\tau_{xy}$ & $0.324 \pm 0.011$ & $0.328 \pm 0.006$ & $0.359 \pm 0.028$ & $0.345 \pm 0.007$ & $0.456 \pm 0.075$ & $0.373 \pm 0.021$\\
    $\tau_{yy}$ & $0.381 \pm 0.043$ & $0.416 \pm 0.013$ & $0.386 \pm 0.055$ & $0.431 \pm 0.009$ & $0.496 \pm 0.110$ & $0.323 \pm 0.105$\\
    $\tau_{zz}$ & $0.210 \pm 0.016$ & $0.240 \pm 0.020$ & $0.238 \pm 0.025$ & $0.235 \pm 0.030$ & $0.469 \pm 0.023$ & $0.406 \pm 0.009$\\
    $F_x$ & $0.595 \pm 0.044$ & $0.653 \pm 0.037$ & $0.679 \pm 0.065$ & $0.686 \pm 0.026$ & $0.746 \pm 0.046$ & $0.722 \pm 0.072$\\
    $F_y$ & $0.512 \pm 0.014$ & $0.489 \pm 0.017$ & $0.494 \pm 0.043$ & $0.504 \pm 0.014$ & $0.555 \pm 0.088$ & $0.460 \pm 0.060$\\
    $C_D$ & $0.136 \pm 0.044$ & $0.176 \pm 0.021$ & $0.152 \pm 0.017$ & $0.168 \pm 0.022$ & $0.279 \pm 0.166$ & $0.191 \pm 0.025$\\
    \bottomrule
  \end{tabular}
  \endgroup
\end{table}

\clearpage
\section{\rev{Sensitivity to the input-filter constant}}\label{app:filter-sensitivity}

\rev{The input filter \eqref{eq:physical-filter} smooths the velocity gradients supplied
to the frozen closure over a length $\delta_f$ that is bounded above by $C_fD$ and decreases
with the local strain rate. To quantify its effect, the in-sample M1 cylinder deployment was
repeated for $C_f\in\{0.05,0.1,0.2,0.3,0.4\}$ with all other settings fixed: the same trained
closure, medium mesh, SUPG treatment, boundary conditions, and the convergence criteria of
\S\ref{sec:method-solver} (relative residuals below $10^{-5}$,
at most 500 Picard iterations). Table~\ref{tab:filter-sensitivity} reports the results.}

\begin{table}
  \centering
  \begingroup\color{black}
  \caption{\rev{Sensitivity of the in-sample M1 cylinder deployment to the input-filter
  constant $C_f$. Errors are relative $L^2$ errors $\mathcal E(q)$ against DNS. The median of
  $\delta_f/D$ is taken over the comparison window. Differences are relative $L^2$ differences
  from the $C_f=0.2$ solution, with the stress difference computed jointly over
  $(\tau_{xx},\tau_{xy},\tau_{yy})$. $C_d$ is a surface control-volume estimate (DNS: $1.21$).}}
  \label{tab:filter-sensitivity}
  \footnotesize
  \setlength{\tabcolsep}{3pt}
  \begin{tabular}{@{}lccccc@{}}
    \toprule
    $C_f$ & $0.05$ & $0.1$ & $0.2$ & $0.3$ & $0.4$\\
    \midrule
    Picard iterations to convergence & 236 & 175 & 117 & 92 & 83\\
    median $\delta_f/D$ & 0.046 & 0.091 & 0.180 & 0.269 & 0.357\\
    \midrule
    $\mathcal E(U)$ & 0.021 & 0.025 & 0.034 & 0.044 & 0.051\\
    $\mathcal E(V)$ & 0.158 & 0.175 & 0.208 & 0.239 & 0.261\\
    $\mathcal E(p)$ & 0.346 & 0.343 & 0.333 & 0.333 & 0.343\\
    $\mathcal E(\tau_{xx})$ & 0.261 & 0.268 & 0.293 & 0.339 & 0.399\\
    $\mathcal E(\tau_{xy})$ & 0.192 & 0.214 & 0.276 & 0.333 & 0.377\\
    $\mathcal E(\tau_{yy})$ & 0.480 & 0.473 & 0.453 & 0.446 & 0.454\\
    $C_d$ & 1.10 & 1.13 & 1.22 & 1.30 & 1.35\\
    \midrule
    velocity difference from $C_f=0.2$ (\%) & 2.31 & 1.60 & --- & 1.32 & 2.22\\
    pressure difference from $C_f=0.2$ (\%) & 9.28 & 6.70 & --- & 6.40 & 11.60\\
    stress difference from $C_f=0.2$ (\%) & 13.89 & 9.76 & --- & 9.54 & 18.21\\
    \bottomrule
  \end{tabular}
  \endgroup
\end{table}

\rev{All five deployments satisfy the convergence criteria, whereas the deployment without
the input filter does not converge within 500 iterations (\S\ref{sec:results-stab}). The
filter constant therefore controls a trade-off between robustness and fidelity. Larger
$C_f$ smooths the closure inputs more strongly and reduces the iteration count from 236 to 83,
but it also removes gradient structure in the shear layers and near wake: the errors in $U$,
$V$, $\tau_{xx}$, and $\tau_{xy}$ increase monotonically with $C_f$, by factors of about
$1.5$--$2.4$ between $C_f=0.05$ and $0.4$. The pressure and $\tau_{yy}$ errors vary by less
than $8\%$, and $C_d$ increases with $C_f$, from $1.10$ to $1.35$. The mean velocity is the
least sensitive quantity, changing by at most $2.3\%$ relative to $C_f=0.2$; the stresses
change by up to $18\%$. These variations exceed the mesh-refinement differences reported in
\S\ref{sec:method-solver}, so $C_f$ is a modelling parameter of the deployed closure rather
than a numerical parameter that vanishes under refinement. We use $C_f=0.2$ throughout as an
intermediate value: it converges in fewer iterations than the weaker filters, remains
within $0.013$ of the most accurate velocity error in this sweep, and gives the drag closest
to DNS. Because $\delta_f$ is independent of the cell size, the same value is used on every
mesh and for all four closures.}

\clearpage

\section{In-sample and leave-one-shape-out field comparisons}\label{app:fields}

\begin{figure}
  \centering
  \includegraphics[width=\textwidth]{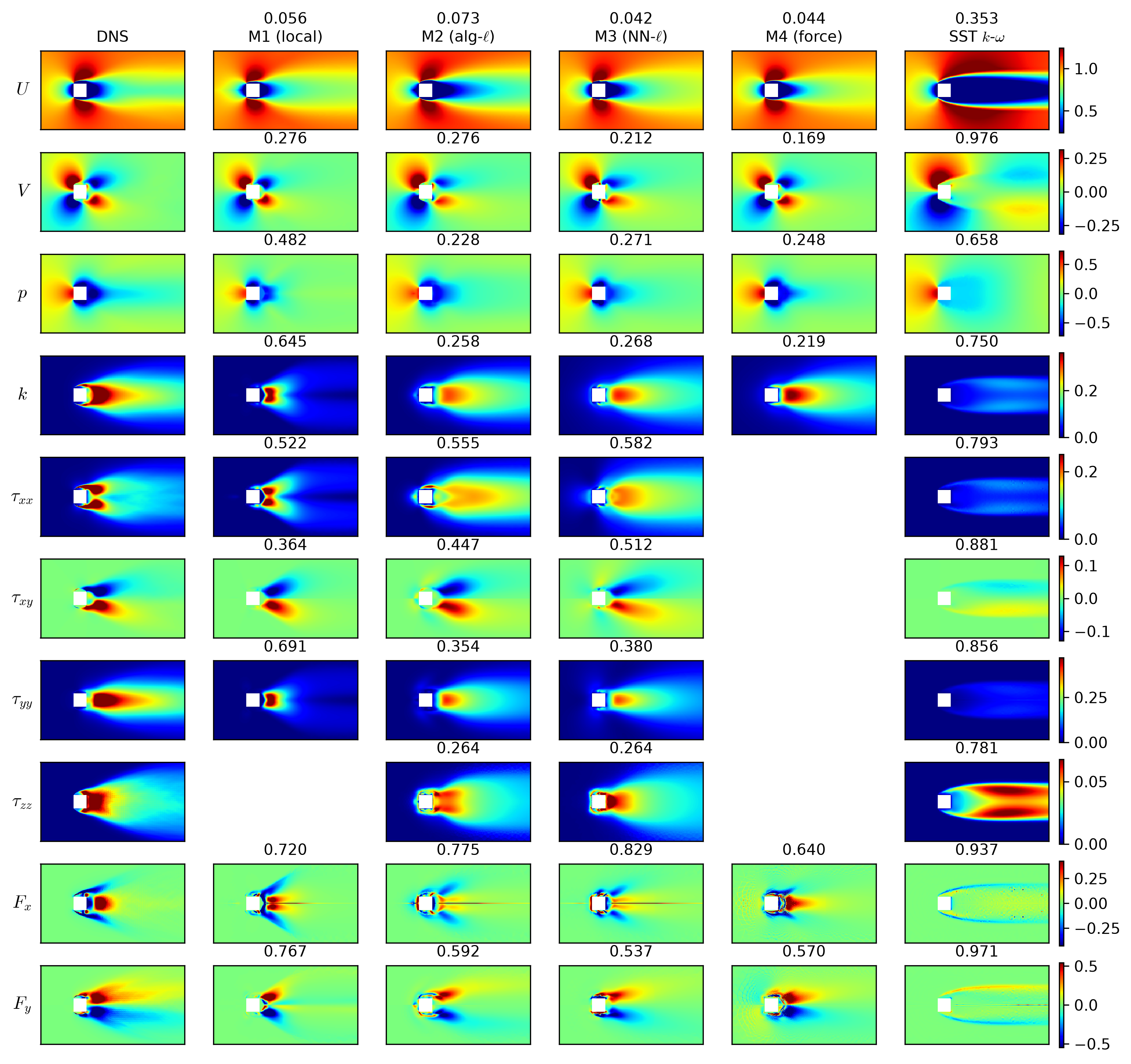}
  \caption{In-sample six-way field comparison for the square. Columns: DNS, M1 (local), M2 (algebraic
  $\lmix$), M3 (learned $\lmix$), M4 (force), SST $k$--$\omega$; rows: $U,V,p,k,\tau_{xx},\tau_{xy},
  \tau_{yy},\tzz,F_x,F_y$; the number above each panel is that field's relative-$L^2$ error against
  DNS (as in figure~\ref{fig:insample}).}
  \label{fig:app-is-sq0}
\end{figure}

\begin{figure}
  \centering
  \includegraphics[width=\textwidth]{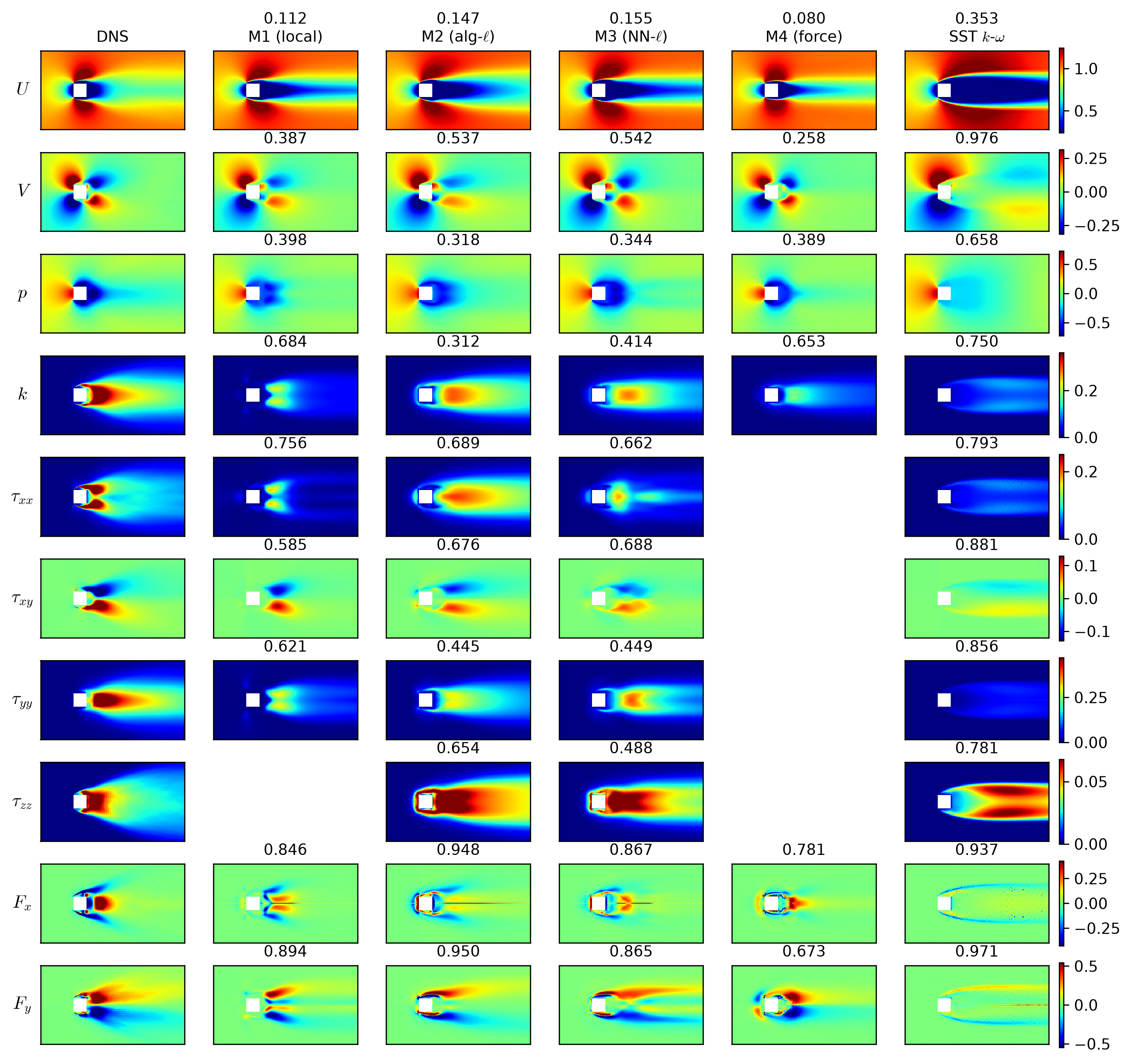}
  \caption{Leave-one-shape-out six-way field comparison for the square (each model trained on the
  other five shapes). Columns: DNS, M1--M4, SST $k$--$\omega$; rows: $U,V,p,k,\tau_{xx},\tau_{xy},
  \tau_{yy},\tzz,F_x,F_y$; the number above each panel is that field's relative-$L^2$ error against
  DNS (as in figure~\ref{fig:insample}).}
  \label{fig:app-lo-sq0}
\end{figure}

\begin{figure}
  \centering
  \includegraphics[width=\textwidth]{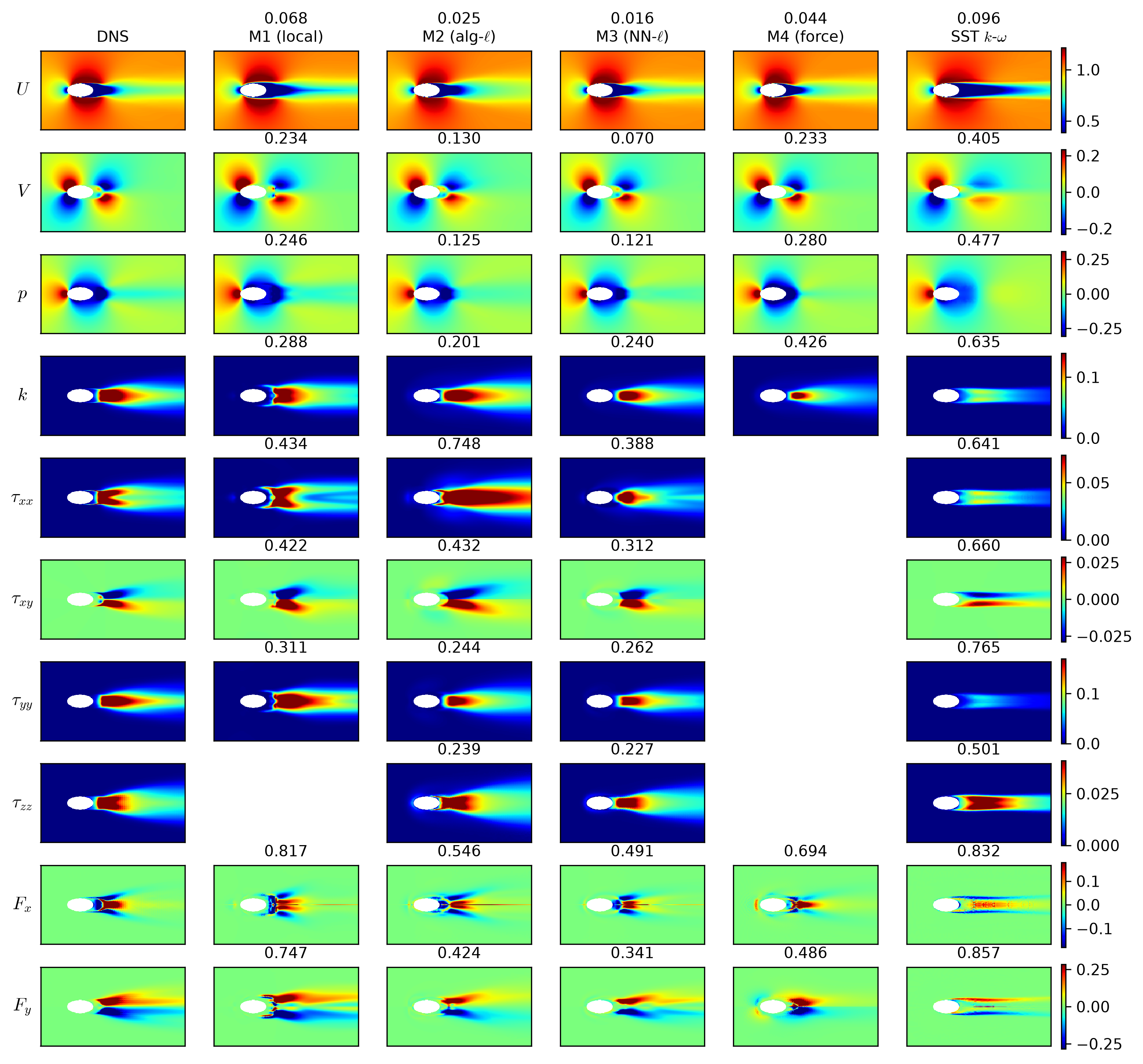}
  \caption{In-sample six-way field comparison for the ellipse. Columns: DNS, M1--M4, SST
  $k$--$\omega$; rows: $U,V,p,k,\tau_{xx},\tau_{xy},\tau_{yy},\tzz,F_x,F_y$; the number above each
  panel is that field's relative-$L^2$ error against DNS (as in figure~\ref{fig:insample}).}
  \label{fig:app-is-ell}
\end{figure}

\begin{figure}
  \centering
  \includegraphics[width=\textwidth]{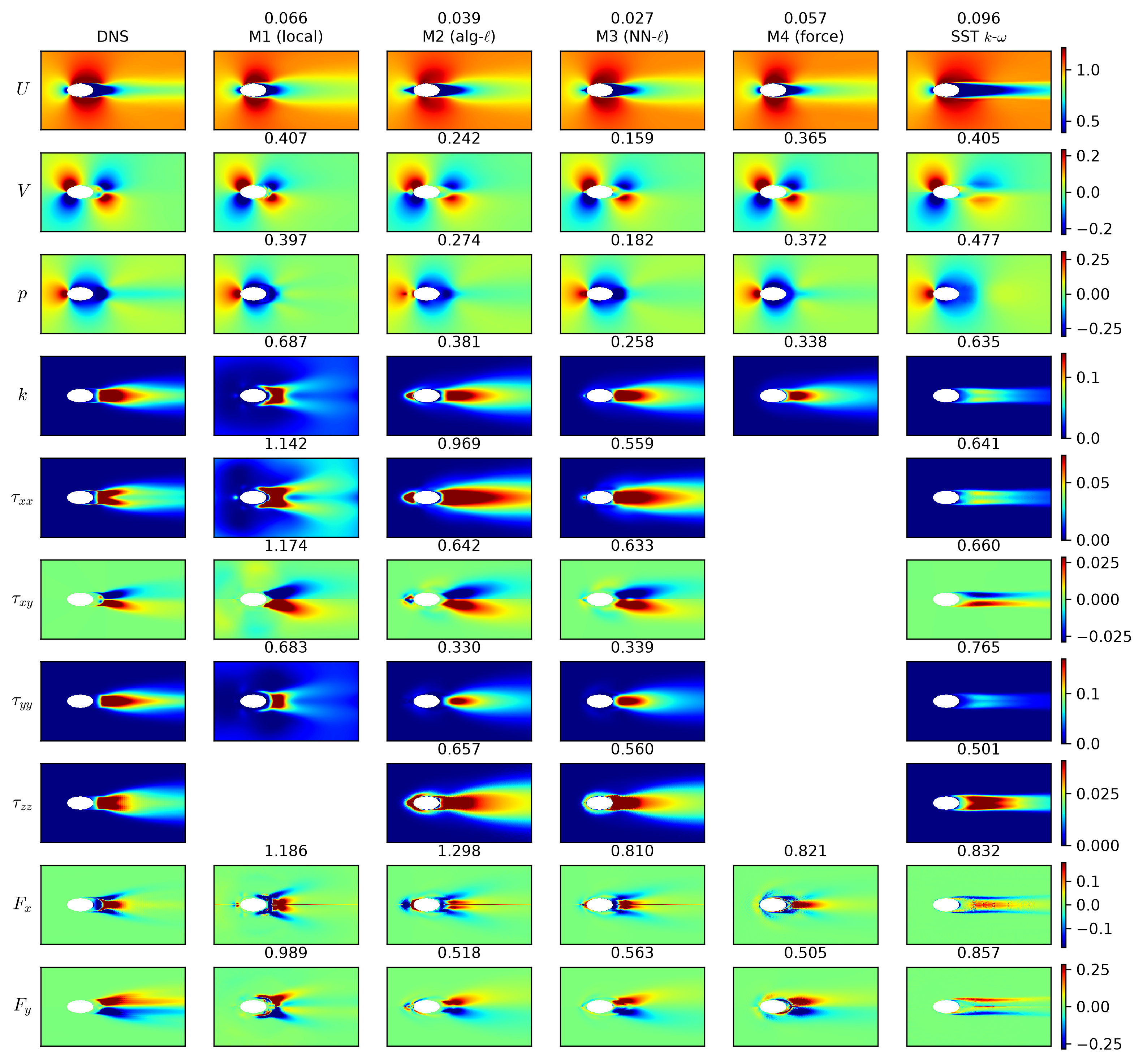}
  \caption{Leave-one-shape-out six-way field comparison for the held-out ellipse---the streamlined,
  smooth-body geometry least like the bluffer training shapes and the hardest case for the stress
  closures. Layout as in figure~\ref{fig:insample}: columns are the DNS reference and the models
  M1--M4 and SST $k$--$\omega$; rows are $U,V,p,k,\tau_{xx},\tau_{xy},\tau_{yy},\tzz,F_x,F_y$; the
  number above each panel is that field's relative-$L^2$ error against DNS.}
  \label{fig:app-lo-ell}
\end{figure}

\begin{figure}
  \centering
  \includegraphics[width=\textwidth]{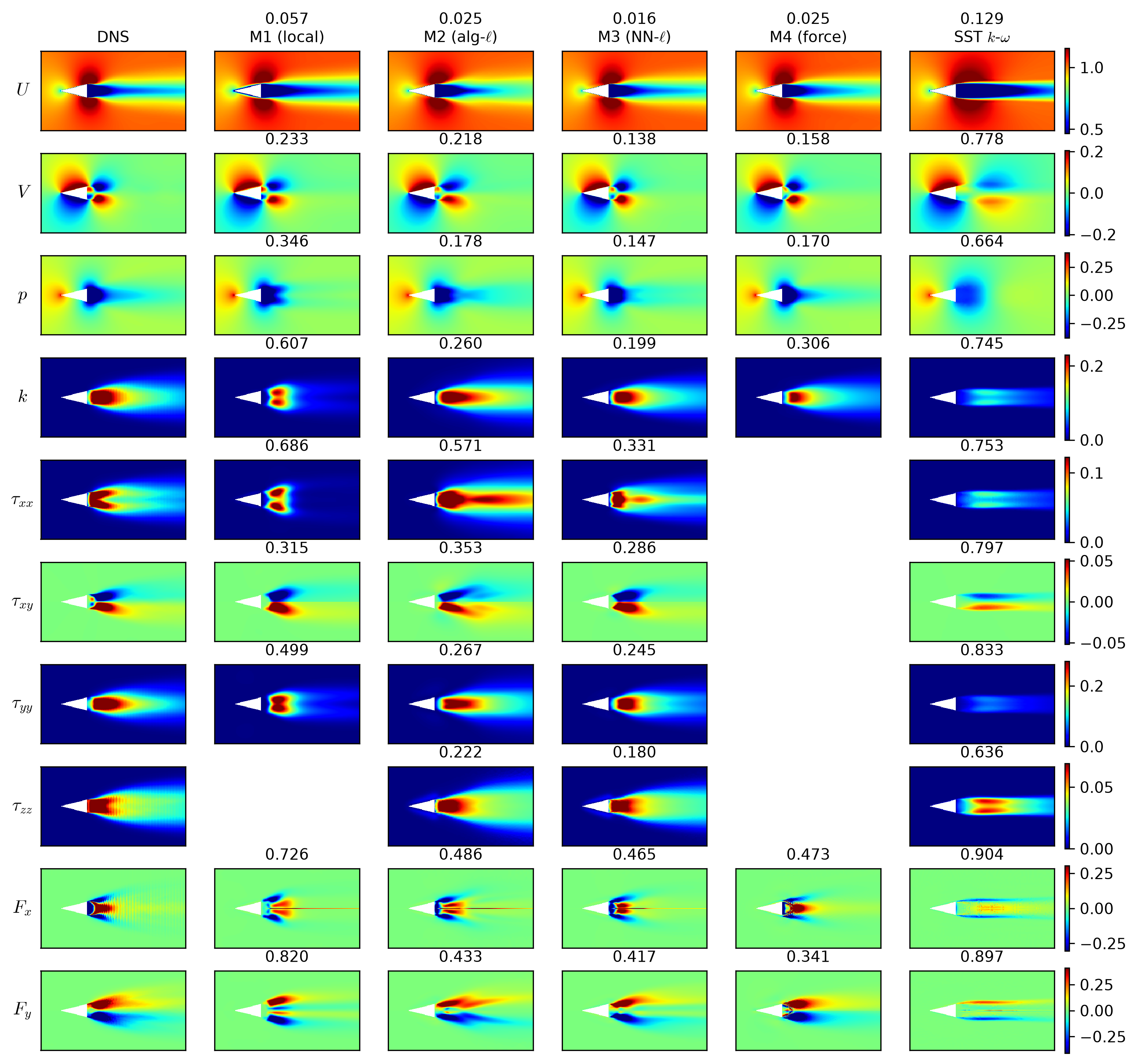}
  \caption{In-sample six-way field comparison for the long triangle. Columns: DNS, M1--M4, SST
  $k$--$\omega$; rows: $U,V,p,k,\tau_{xx},\tau_{xy},\tau_{yy},\tzz,F_x,F_y$; the number above each
  panel is that field's relative-$L^2$ error against DNS (as in figure~\ref{fig:insample}).}
  \label{fig:app-is-trilong}
\end{figure}

\begin{figure}
  \centering
  \includegraphics[width=\textwidth]{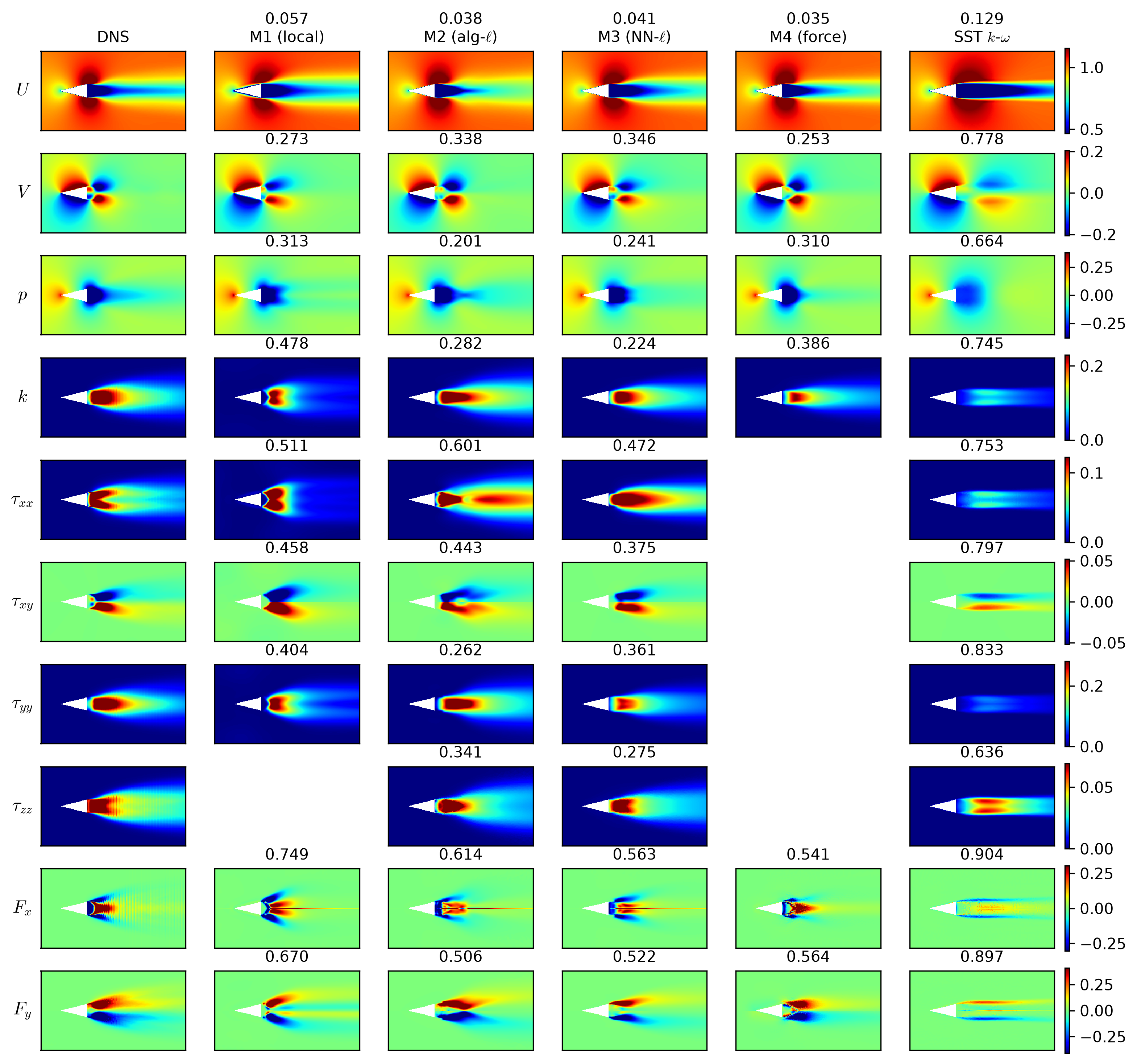}
  \caption{Leave-one-shape-out six-way field comparison for the long triangle (each model trained on
  the other five shapes). Columns: DNS, M1--M4, SST $k$--$\omega$; rows: $U,V,p,k,\tau_{xx},
  \tau_{xy},\tau_{yy},\tzz,F_x,F_y$; the number above each panel is that field's relative-$L^2$ error
  against DNS (as in figure~\ref{fig:insample}).}
  \label{fig:app-lo-trilong}
\end{figure}

\begin{figure}
  \centering
  \includegraphics[width=\textwidth]{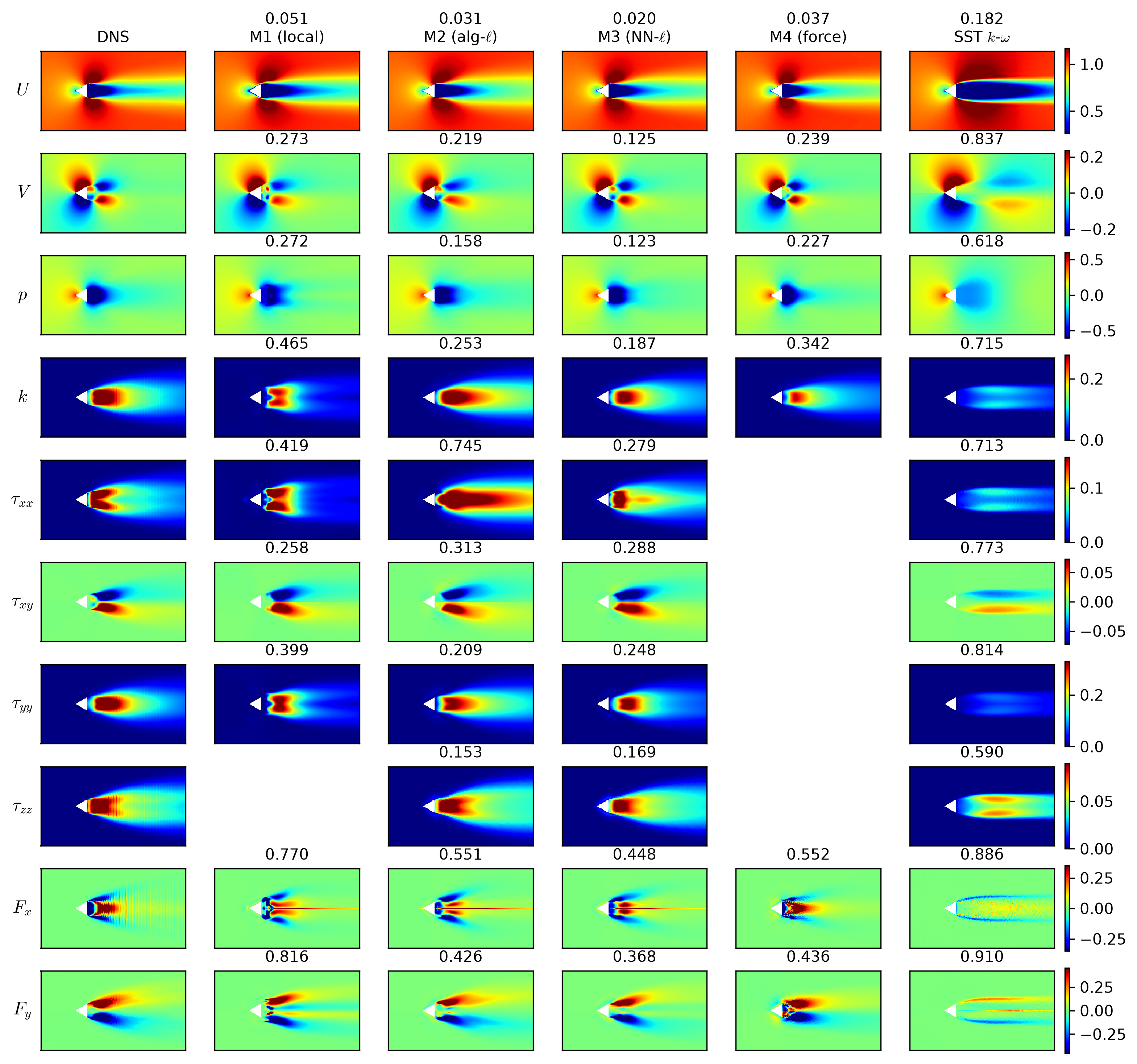}
  \caption{In-sample six-way field comparison for the equilateral triangle. Columns: DNS, M1--M4, SST
  $k$--$\omega$; rows: $U,V,p,k,\tau_{xx},\tau_{xy},\tau_{yy},\tzz,F_x,F_y$; the number above each
  panel is that field's relative-$L^2$ error against DNS (as in figure~\ref{fig:insample}).}
  \label{fig:app-is-triequi}
\end{figure}

\begin{figure}
  \centering
  \includegraphics[width=\textwidth]{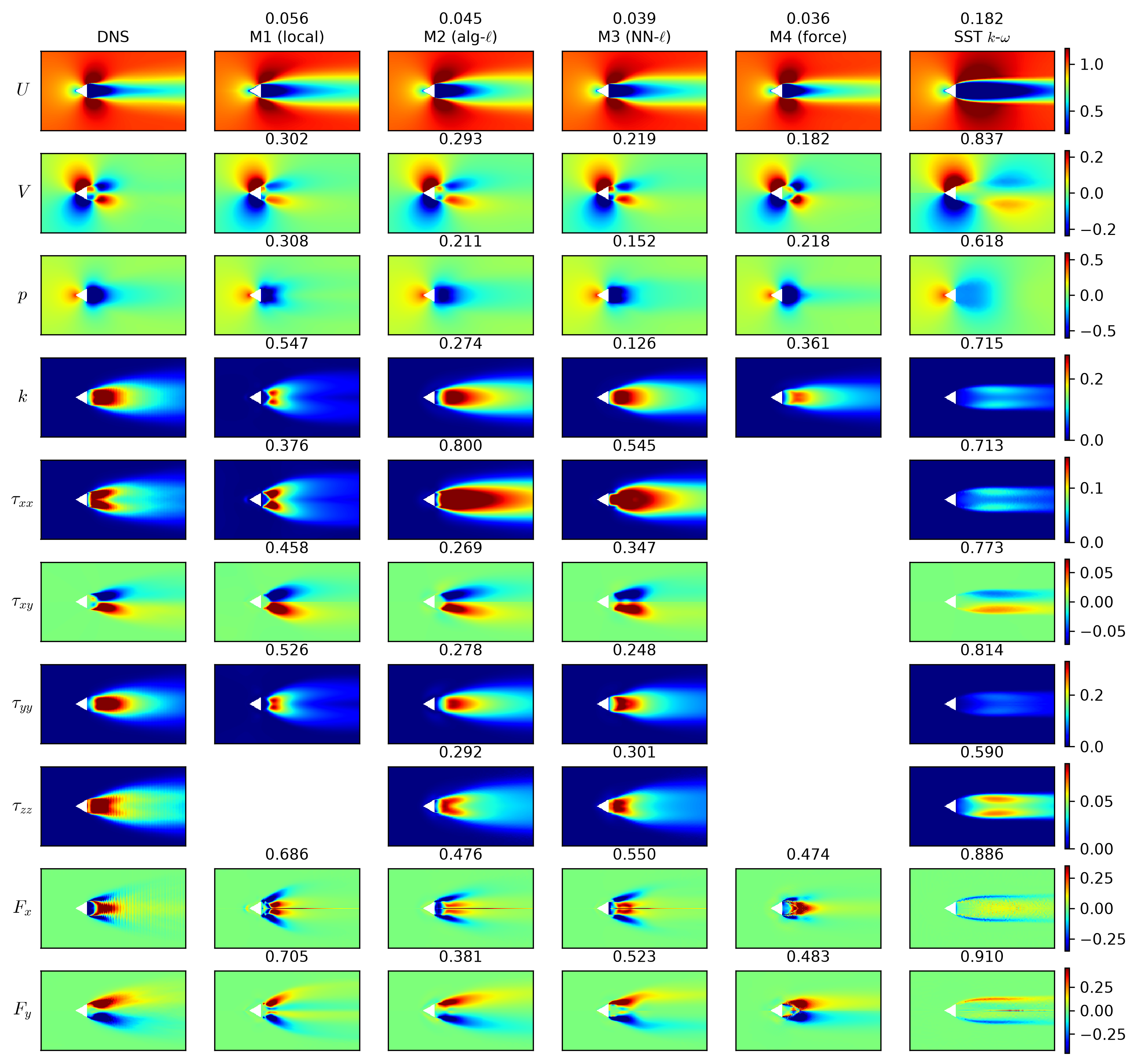}
  \caption{Leave-one-shape-out six-way field comparison for the equilateral triangle (each model
  trained on the other five shapes). Columns: DNS, M1--M4, SST $k$--$\omega$; rows: $U,V,p,k,
  \tau_{xx},\tau_{xy},\tau_{yy},\tzz,F_x,F_y$; the number above each panel is that field's
  relative-$L^2$ error against DNS (as in figure~\ref{fig:insample}).}
  \label{fig:app-lo-triequi}
\end{figure}

The six-way field comparisons (DNS and the five deployed models M1--M4 and SST) for the circular
cylinder appear in the main text, in-sample and leave-one-shape-out (figures~\ref{fig:insample}
and~\ref{fig:loso}), as do the diamond (figures~\ref{fig:app-is-sq45},~\ref{fig:app-lo-sq45}).
The remaining comparisons are collected
here \rev{as in-sample and leave-one-shape-out pairs for each shape} (figures~\ref{fig:app-is-sq0}--\ref{fig:app-lo-triequi}). In every panel the rows are $U$, $V$, $p$,
$k$, $\tau_{xx}$, $\tau_{xy}$, $\tau_{yy}$, $\tzz$, and the Reynolds-force components $F_x$, $F_y$;
the columns are DNS, M1 (local), M2 (algebraic $\lmix$), M3 (learned $\lmix$), M4 (force), and SST
$k$--$\omega$; the number above each panel is that field's relative-$L^2$ error against DNS (smaller
is better, the DNS column carrying none); and blank panels are quantities a model does
not represent (M1 omits $\tzz$ and the force model M4 omits the four stresses; the force $F_x,F_y$ of
the stress models M1--M3 is the divergence of their predicted stress), exactly as in
figure~\ref{fig:insample}.

\rev{\emph{Square} (figures~\ref{fig:app-is-sq0},~\ref{fig:app-lo-sq0}). In-sample, all four
closures reduce the mean-velocity error from $\mathcal E(U)=0.353$ for SST to $0.042$--$0.073$
and the cross-stream error from $\mathcal E(V)=0.976$ to $0.169$--$0.276$. The transported-$k$
closures M2--M4 also reduce the energy error from $0.750$ to $0.219$--$0.268$, and M2 and M3
reduce the cross-stream normal stress error from $0.856$ to $0.354$ and $0.380$. The force model
M4 is the most accurate on $V$, $k$, and $F_x$. The held-out square is the hardest case for the
mean velocity: $\mathcal E(U)$ rises to $0.080$--$0.155$, and $\mathcal E(V)$ of M2 and M3 to
$0.54$. M4 degrades least ($\mathcal E(U)=0.080$, $\mathcal E(V)=0.258$) and gives the most
accurate Reynolds force ($\mathcal E(F_x)=0.781$, $\mathcal E(F_y)=0.673$). All closures remain
well below the SST errors on the mean velocity and pressure.

\emph{Ellipse} (figure~\ref{fig:app-is-ell}). The streamlined ellipse is the case in which SST
is most accurate ($\mathcal E(U)=0.096$), but the learned closures still improve on it
substantially. M3 reduces the errors in $U$, $V$, and $p$ to $0.016$, $0.070$, and $0.121$,
four to six times below SST, and halves the shear-stress error ($0.312$ against $0.660$). The
algebraic length scale of M2 again over-predicts the streamwise normal stress
($\mathcal E(\tau_{xx})=0.748$ against $0.641$ for SST), which the learned length scale of M3
corrects ($0.388$). The held-out ellipse \rev{(figure~\ref{fig:app-lo-ell})} is discussed in \S\ref{sec:results-loso}.

\emph{Long triangle} (figures~\ref{fig:app-is-trilong},~\ref{fig:app-lo-trilong}). In-sample,
M3 is the most accurate closure on nearly every field, with $\mathcal E(U)=0.016$,
$\mathcal E(V)=0.138$, $\mathcal E(p)=0.147$, $\mathcal E(k)=0.199$, and
$\mathcal E(\tau_{xx})=0.331$, against $0.129$, $0.778$, $0.664$, $0.745$, and $0.753$ for SST.
This shape generalizes well: for the held-out long triangle the errors increase only
moderately, M4 recovers the mean velocity best ($\mathcal E(U)=0.035$, $\mathcal E(V)=0.253$),
M3 the energy and stresses ($\mathcal E(k)=0.224$, $\mathcal E(\tau_{xx})=0.472$,
$\mathcal E(\tau_{xy})=0.375$), and every closure is below SST on every field.

\emph{Equilateral triangle} (figures~\ref{fig:app-is-triequi},~\ref{fig:app-lo-triequi}).
In-sample, M3 again gives the lowest errors on the mean flow, energy, and streamwise normal stress
($\mathcal E(U)=0.020$, $\mathcal E(V)=0.125$, $\mathcal E(p)=0.123$, $\mathcal E(k)=0.187$,
$\mathcal E(\tau_{xx})=0.279$), while M2 shows the same over-prediction of $\tau_{xx}$ as for the
ellipse. On the held-out equilateral triangle, M4 is the most accurate on the mean velocity
($\mathcal E(U)=0.036$, $\mathcal E(V)=0.182$) and M3 on the pressure and energy
($\mathcal E(p)=0.152$, $\mathcal E(k)=0.126$), against $0.618$ and $0.715$ for SST.

Across these seven comparisons, the learned closures are more accurate than SST for almost every
field. The exceptions are the streamwise normal stress of M2 on the ellipse and the equilateral
triangle, a consequence of its algebraic length scale, and the streamwise Reynolds force of M2 on
the held-out square ($0.948$ against $0.937$). The largest and most consistent improvements are
in the mean velocity, the pressure, and the turbulent kinetic energy; for the mean velocity, the
error of the best closure is six to nine times smaller than that of SST in-sample.}

\clearpage

\clearpage  

\end{document}